\documentclass[usenatbib]{mn2e}
\usepackage{epsfig}
\usepackage{amsmath}
\usepackage{ulem}
\usepackage{times}
\usepackage{longtable}
\usepackage{supertabular}
\usepackage{color}

\def\gsim{\mathrel{\raise0.35ex\hbox{$\scriptstyle >$}\kern-0.6em 
\lower0.40ex\hbox{{$\scriptstyle \sim$}}}}
\def\lsim{\mathrel{\raise0.35ex\hbox{$\scriptstyle <$}\kern-0.6em 
\lower0.40ex\hbox{{$\scriptstyle \sim$}}}}

\def\oii{{\rm [O{\sc ii}]}}
\def\oiii{{\rm [O{\sc iii}]}}

\date{\today}
\title[Massive starbursts in a $z\sim 2$ proto-cluster]
{Massive starburst galaxies in a {\boldmath $z=2.16$} proto-cluster unveiled \\
by panoramic H{\boldmath $\alpha$} mapping}

\author[Y. Koyama et al.]{
\parbox[t]{\textwidth}{
Yusei Koyama,$^{\! 1,2}$\thanks{E-mail: yusei.koyama@durham.ac.uk}
Tadayuki Kodama,$^{\! 1,3}$
Ken-ichi Tadaki,$^{\! 4}$
Masao Hayashi,$^{\! 1}$\\
Masayuki Tanaka,$^{\! 5}$
Ian Smail,$^{\! 6}$
Ichi Tanaka,$^{\! 1,3}$
Jaron Kurk$^{7}$
}
\vspace*{6pt}\\
$^{1}$National Astronomical Observatory of Japan, Mitaka, Tokyo
181-8588, Japan\\
$^{2}$Department of Physics, Durham University, South Road, Durham DH1 3LE, UK\\
$^{3}$Subaru Telescope, National Astronomical Observatory of Japan, 650 
North A'ohoku Place, Hilo, HI 96720, USA\\
$^{4}$Department of Astronomy, Graduate School of Science, The University of Tokyo, Tokyo 113-0033, Japan \\
$^{5}$Institute for the Physics and Mathematics of the Universe, The University of Tokyo, 5-1-5 Kashiwanoha, Kashiwa-shi, Chiba 277-8583, Japan\\
$^{6}$Institute for Computational Cosmology, Durham University, South Road, Durham DH1 3LE, UK\\
$^{7}$Max-Planck-Institut f{\"u}r Extraterrestrische Physik, Postfach 1312, Giessenbachstrasse, D-85741 Garching, Germany
}
\begin{document}

\maketitle

%------------------------------------------------------------------
\begin{abstract}

We present a panoramic narrow-band study of H$\alpha$ emitters in the field of the $z=2.16$ proto-cluster around PKS\,1138$-$262 using MOIRCS on the Subaru Telescope. We find 83 H$\alpha$ emitters down to a SFR\,(H$\alpha$)\,$\sim$\,10$M_{\odot}$\,yr$^{-1}$ across a $\sim$7$'\times$7$'$ region centered on the radio galaxy, and identify $\sim$\,10-Mpc scale filaments of emitters running across this region. By examining the properties of H$\alpha$ emitters within the large-scale structure, we find that galaxies in the higher-density environments at $z=2.16$ tend to have redder colours and higher stellar masses compared to galaxies in more underdense regions. We also find a population of H$\alpha$ emitters with red colours ($(J-K_s)\gsim 1$), which are much more frequent in the denser environments and which have apparently very high stellar masses with $M_*\gsim$\,10$^{11}M_{\odot}$, implying that these cluster galaxies have already formed a large part of their stellar mass before $z\sim 2$. {\it Spitzer Space Telescope} 24$\mu$m data suggests that many of these red H$\alpha$ emitters are bright, dusty starbursts (rather than quiescent sources). We also find that the proto-cluster galaxies follow the same correlation between SFR and $M_*$ (the ``main sequence'') of $z\sim 2$ field star-forming galaxies, but with an excess of massive galaxies. These very massive star-forming galaxies are not seen in our similar, previous study of $z\sim 1$ clusters, suggesting that their star-formation activity has been shut off at $1\lsim z \lsim2$. We infer that the massive red (but active) galaxies in this rich proto-cluster are likely to be the products of environmental effects, and they represent the accelerated galaxy formation and evolution in a biased high density region in the early Universe.

\end{abstract}
%------------------------------------------------------------------
%%%%%%%%%%%%%%%%%%%%%%%%%%%%%%%%%%%%%%%%%%%%%%%%%%%%%%%%%%%%%%%%%%%%%%%%%%
% INTRODUCTION
%%%%%%%%%%%%%%%%%%%%%%%%%%%%%%%%%%%%%%%%%%%%%%%%%%%%%%%%%%%%%%%%%%%%%%%%%%
\section{Introduction}
\label{sec:intro}

It is widely recognized that galaxy properties such as star-formation rate (SFR) and morphology depend  on the environment in which the galaxies reside (e.g. \citealt{dre80}; \citealt{kod01}; \citealt{lew02}; \citealt{gom03}; \citealt{got03}; \citealt{tan04}). Local galaxy clusters are dominated by red quiescent ellipticals and S0s, and their tight colour--magnitude relation suggests very old stellar populations in these cluster galaxies (e.g. \citealt{bow92}; \citealt{kod98}; \citealt{smi12}). This means that many present-day cluster galaxies must have formed their stars in the early Universe at $z\gg 1$, and therefore distant clusters of galaxies are an ideal site for studying this early evolutionary stage of cluster galaxies. 

Earlier studies have identified a high fraction of star-forming population in distant clusters (e.g. \citealt{but84}).  In addition to this blue star-forming population, optically red star-forming galaxies, as well as a starbursting population revealed by mid-infrared (MIR) and radio observations are also reported to increase in frequency in distant clusters (e.g.\ \citealt{sma99}; \citealt{gea06}; \citealt{mar07}; \citealt{sai08}; \citealt{hai09}), particularly in the cluster outskirts (e.g.\ \citealt{koy08}; 2011). In the very rich cluster cores, however, the star-forming galaxy fraction is still very low up to $z\sim 1$ (e.g.\ \citealt{cou01}; \citealt{bal02}; \citealt{kod04}; \citealt{koy10}; \citealt{sob11}; \citealt{bau11}), suggesting that the major phase of star formation for galaxies in these regions must be occurring at an even earlier epoch. Indeed some recent studies have also suggested a ``reversal'' of star formation--density relation at $z\sim 1$ (e.g.\ \citealt{elb07}; \citealt{coo08})  in high-density environments in the distant Universe. Therefore, the  next challenge must be to study directly the site of cluster galaxy formation at $z\gg 1$. New surveys have unveiled several clusters of galaxies at these early times, $z\gsim 1.5$, using a variety of techniques (e.g.\ \citealt{sta06}; \citealt{and09}; \citealt{tan10a}; \citealt{pap10}; \citealt{gob11}), and these are good laboratories for studying the early activity of cluster galaxies (e.g.\ \citealt{qua12}; \citealt{rai12}). By conducting an [O{\sc ii}] emission-line survey of a $z=1.46$ cluster with the Subaru Telescope, \cite{hay10} first noted a large fraction of [O{\sc ii}] galaxies in the cluster, and they highlighted the high activity of cluster core galaxies at $z\sim 1.5$. Subsequent studies of similarly high-redshift clusters also confirm active galaxy populations in these high-density environments (\citealt{tra10}; \citealt{hil10}; \citealt{fas11}; \citealt{tad12}), suggesting we are approaching the formation epoch of present-day cluster galaxies. 

It is still challenging to detect over-dense regions in the more distant Universe at $z\gsim 2$ (i.e.\ the peak epoch of galaxy activity in the general field population; e.g.\ \citealt{hop06}). Bright high-redshift radio galaxies or quasors (QSOs), or indeed overdensities comprising several QSOs, have been used as landmarks of high-redshift over-dense regions. This approach has been successful in the last decade, and many ``proto-clusters'' have been identified (e.g.\ \citealt{ove06}; \citealt{ven07}; \citealt{mil08} and references therein). One of the most effective ways to select ``member'' galaxies associated to the proto-clusters is to exploit narrow-band filters to pick up strong emission-lines (such as Ly$\alpha$, [O{\sc ii}] or H$\alpha$) at the particular redshifts (e.g.\ \citealt{ste00}; \citealt{kur00}; 2004a; \citealt{mat11}; \citealt{yam12}). Considering the importance of dust-obscured activity in the early Universe as stated above, the H$\alpha$ line is the preferred tracer because it is less affected by dust extinction (or metallicity) compared to other strong lines emitted in the rest-frame optical or ultra-violet (UV), which tend to be biased to lower-metallicity galaxies with little dust extinction. An observational challenge is that the H$\alpha$ lines of distant galaxies (at $z\gsim 0.5$) shift to near-infrared regime (NIR), where the night sky emission is strong, and where efficient large format of detectors became available only recently. 

The PKS\,1138$-$262 field is one of the best-studied proto-clusters, centered on a radio galaxy at $z=2.156$, and importantly, the H$\alpha$ line from this redshift falls in between the night sky emission lines in the NIR and can be observed with ground-based telescopes (\citealt{kur04a}b). The central radio galaxy (PKS\,1138) is one of the brightest radio galaxies known at $z\gsim 2$ with a clumpy morphology in the optical (\citealt{pen97}; 1998) and distorted morphology in radio (\citealt{car97}; 2002). The pioneering Ly$\alpha$ imaging and spectroscopy around this radio galaxy was performed by \cite{kur00} and \cite{pen00}. Those studies spectroscopically confirmed 14 Ly$\alpha$ emitting galaxies at $z\simeq 2.16$. \cite{pen02} used X-ray observations to identify 18 X-ray sources in the PKS\,1138 field, among which five were confirmed to be cluster members by subsequent spectroscopic campaign of this field (\citealt{kur04b}; \citealt{cro05}). An intensive search for H$\alpha$ emitters and extremely red objects (EROs) was also performed by \cite{kur04a} with VLT. They reported $\sim$\,40 H$\alpha$ emitter candidates within a $\sim 12$\,arcmin$^{2}$ field, 9 of them being spectroscopically confirmed (3 have detectable H$\alpha$+[N{\sc ii}] emission in their NIR spectra; see \citealt{kur04b} for details). Following these studies, more recent work has also found further evidence that the PKS\,1138 is a forming proto-cluster (\citealt{mil06}; \citealt{kod07}; \citealt{zir08}; \citealt{hat08}; 2009; \citealt{doh10}; \citealt{tan10b}; \citealt{kui11}). Given the strong evidence of the over-density of galaxies, this PKS\,1138 field may be an ideal laboratory for testing the environmental dependence of galaxy properties at $z\sim 2$ (\citealt{tan10b}). 

In this paper, we revisit the PKS\,1138 field using observations from the wide-field NIR camera, MOIRCS (\citealt{ich06}; \citealt{suz08}) on the Subaru Telescope (\citealt{iye04}). Our aim is to map the entire structure around this proto-cluster and the environmental dependence of H$\alpha$-based star forming activity at $z\sim 2$ within this structure.  We show our Subaru data and supplementary data in \S2. The H$\alpha$ emitter selection procedure and the derivation of physical quantities are presented in \S3. Our main results and discussions are shown in \S4, and we give a summary of this paper in \S5. Throughout this paper, we adopt the standard cosmology with $\Omega_M =0.3$, $\Omega_{\Lambda} =0.7$, and $H_0 =70$\,km\,s$^{-1}$\,Mpc$^{-1}$, which gives a 1$''$ scale of 8.29 kpc at the redshift of the PKS\,1138: $z=2.156$. Magnitudes are given in the AB system, unless otherwise stated.

%%%%%%%%%%%%%%%%%%%%%%%%%%%%%%%%%%%%%%%%%%%%%%%%%%%%%%%%%%%%%%%%%%%%%%%%%%
% DATA
%%%%%%%%%%%%%%%%%%%%%%%%%%%%%%%%%%%%%%%%%%%%%%%%%%%%%%%%%%%%%%%%%%%%%%%%%%
\section{Data}
\label{sec:data}

\subsection{MOIRCS data}

\begin{figure}
  \begin{center}
    \leavevmode
    \vspace{-1.2cm}
    \rotatebox{0}{\includegraphics[width=8.5cm,height=8.5cm]{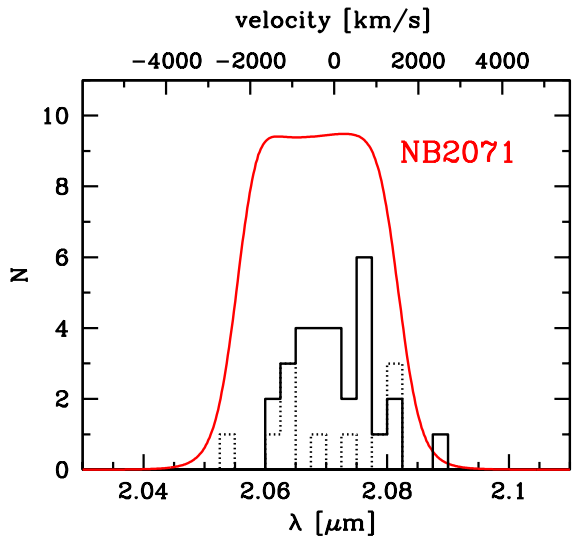}}
  \end{center}
  \vspace{-0.5cm}
  \caption{ The velocity distribution (relative to the radio galaxy) of the spectroscopically confirmed proto-cluster member galaxies shown by Pentericci et al.\ (2000), Kurk et al.\ (2004b), and Doherty et al.\ (2010) (solid-line histogram), with an arbitrarily scaled transmission curve of the NB2071 filter. We also show the velocity distribution of the emission-line sources close to the radio galaxy (within $\sim$15$''\times$15$''$) revealed with integral field spectroscopy by Kuiper et al.\ (2011) (dotted-line histogram). 
}
\label{fig:NB2071_trans}
 \end{figure}

We are undertaking a long-term study of galaxies in dense environments at high redshifts: {\it MApping HAlpha and Lines of Oxygen with Subaru}  (MAHALO-Subaru; see overview by \citealt{kod12}). As a part of this project, we observed the field around the $z=2.156$ radio galaxy PKS\,1138 through the $J$, $K_s$, and NB2071 filters using the MOIRCS imaging spectrograph on the Subaru Telescope that has a 4$'\times$7$'$ field. The NB2071 filter ($\lambda_c=2.068$$\mu$m, $\Delta\lambda=0.027$$\mu$m) covers the H$\alpha\lambda$6563 line at $z\simeq 2.13$--2.17, corresponding to velocities of $-2500 \lsim \Delta v \lsim 1500$\,km\,s$^{-1}$ relative to the radio galaxy (see Fig.~\ref{fig:NB2071_trans}). Note that the NB2071 filter was fabricated for these specific observations, and therefore this filter {\it perfectly} matches to the redshift distribution of spectroscopically confirmed member galaxies in the structure around PKS\,1138 (see Fig.~\ref{fig:NB2071_trans}). Based on the known redshifts (\citealt{kur04b}), we expect to be able to detect $>$\,90\% of the H$\alpha$ emitting cluster members with this filter.

We observe two pointings with MOIRCS, PKS\,1138-C and PKS\,1138-S (each 4$'\times$7$'$ in extent). The PKS\,1138-C field (R.A.\ 11$^{\rm h}$40$^{\rm m}$48$^{\rm s}$.4, Dec.\ $-$26$^{\rm d}$29$^{\rm m}$09$^{\rm s}$.0, J2000) is centered on the radio galaxy, and the PKS\,1138-S field centre (R.A.\ 11$^{\rm h}$40$^{\rm m}$48$^{\rm s}$.4, Dec.\ $-$26$^{\rm d}$32$^{\rm m}$44$^{\rm s}$.0, J2000) is $\sim$\,3.5$'$ away from the radio galaxy towards the south. Note that this H$\alpha$ emitter survey covers a factor of $\sim$\,4\,$\times$ larger area than the previous H$\alpha$ survey with ISAAC on the VLT by \cite{kur04a}, and this is indeed one of the widest-field H$\alpha$ survey targeting a high-redshift structure to date. The observation were carried out under good conditions (see the data summary in Table~1), and the data were reduced in a standard manner with {\sc mcsred} \footnote{http://www.naoj.org/staff/ichi/MCSRED/mcsred\_e.html} software, which was written by one of the co-authors (I.\ Tanaka; see \citealt{tani11}). We use the {\sc nbmcsall} package in {\sc mcsred} for narrow-band (NB) data reduction, which was designed to reduce MOIRCS NB data. We also use the $J$- and $K_s$-band imaging data of  the PKS\,1138-C field obtained in 2006 presented by \cite{kod07}. We mosaiced all the data together, and the final image size is 6.7$'\times$\,7.5$'$.
Note that all the data are smoothed to 0.7$''$ FWHM seeing before measuring the multi-band photometry, and all the analyses in this paper were performed on these smoothed images. We estimated the limiting magnitudes of each image by measuring the variance of 1.5$''$ diameter aperture photometry at random positions on each image. The data quality is effectively uniform, but the limiting magnitudes  differ slightly between the PKS\,1138-C and PKS\,1138-S fields, due to the small differences in their total exposure times (see Table~1).  

%------------------------------------------------------------------------
% Fig. - filter transmission
%------------------------------------------------------------------------
%------------------------------------------------------------------------

%-------------------------------------------------------------------
% Optical data
%-------------------------------------------------------------------
\subsection{Optical data}

We obtained a new deep $z'$-band image with Suprime-Cam on the Subaru Telescope. This observation was executed on 29th April 2011 under photometric conditions (see Table~1). We reduced the data in a standard manner with the {\sc sdfred2} software (\citealt{yag02}; \citealt{ouc04}). Note that the field of view of Suprime-Cam (34$'\times$\,27$'$) is much larger than our MOIRCS data coverage, and so we trimmed the $z'$-band image accordingly. The limiting magnitude was estimated in the same way as for the MOIRCS data. 

%-------------------------------------------------------------------
\begin{table*}
\caption{A summary of our Subaru imaging data for the PKS\,1138 fields. 
The PSFs are measured on the mosaiced images (they are finally 
smoothed to 0.7$''$ FWHM for multi-band photometry). Note that we re-use the 
$J$ and $K_s$ data taken in 2006 which were presented in 
Kodama et al.\ (2007). }
\begin{tabular}{l|c|c|c|c|c|c|c}
\hline
\hline
Field  & Filter  &  Instrument  &  FoV  &  Obs. date &  Exp. time  & Limit mag. &  PSF  \\
  &   &    &    &    &  [min]  & (1.5$''$, 5$\sigma$) & FWHM   \\
\hline
PKS\,1138  &  $z'$ & Suprime-Cam &  27$' \times$ 34$'$   &  2011/4/29  &
75   &  25.8  &  0.7$''$  \\
PKS\,1138-C  &  $J$ & MOIRCS &  4$' \times$ 7$'$  & 2006/1/7, 2011/4/14 &  151  &  24.5  &  0.7$''$  \\
PKS\,1138-S  &  $J$  &  MOIRCS &  4$' \times$ 7$'$    &  2011/4/17  &  78  &  24.2  & 0.5$''$   \\
PKS\,1138-C  &  $K_s$ & MOIRCS &  4$' \times$ 7$'$   & 2006/1/6  &  55  &  23.2  &  0.7$''$   \\
PKS\,1138-S  &  $K_s$  & MOIRCS &  4$' \times$ 7$'$  & 2011/4/14, 2011/4/17 & 36  &  23.2  & 0.5$''$   \\
PKS\,1138-C  &  NB2071 & MOIRCS &  4$' \times$ 7$'$ & 2011/3/11  &  186  &  22.9  &  0.7$''$   \\
PKS\,1138-S  &  NB2071  & MOIRCS &  4$' \times$ 7$'$  & 2011/4/14,17 & 119 &  22.7 & 0.5$''$   \\
\hline 
\end{tabular}
\label{tab:data_summary}
\end{table*}
%------------------------------------------------------------------

We also use the $B$-band data taken with FORS1 on VLT by \cite{kur00}.
Note that the $B$-band data and our MOIRCS data do not completely overlap
with each other. In addition the exposure time of this $B$-band data is 
only 30 minutes. As a result the limiting magnitude of the $B$-band data, 
estimated in the same manner as for the Subaru data, is 26.2 mag which 
is not deep enough for studying passive $z\sim 2$ galaxies. But we can 
still use the $B$-band data to check the {\it BzK}  diagram for narrow-band 
sources in \S3.2. The seeing measured from the $B$-band image is 
$\sim$\,0.7$''$ (see \citealt{kur04a}).

\subsection{Photometric catalogue}

We construct a photometric catalogue of our data using the {\sc SExtractor} software package (\citealt{ber96}). The source extraction is based on the NB2071 image, and we use the double-image mode of {\sc SExtractor} to perform multi-band photometry. After removing saturated sources, the catalogue includes 1038 sources with $>5\sigma$ detection in NB2071 (i.e.\ $m_{\textrm{NB2071}}<22.9$). We use {\sc mag\_aper} (with 1.5$''$ diameter apertures) for source detection and measuring colours, and {\sc mag\_auto} for deriving the physical quantities such as star formation rates. Photometric zero points are derived from standard star observation of GD71 for NB2071 and GD153 for the $z'$-band, while we scale the $J$ and $K_s$ data to match the existing data presented in \cite{kod07}, adopting their updated zero points \footnote{The photometric zero-points used in \cite{kod07} were not accurate, and we have corrected them by the following amounts in the current paper; $J_{\rm corr}=J-0.25$ and $K_{s,{\rm corr}}=K_s-0.30$.} (see also \citealt{tan10b}; \citealt{doh10}). We here apply a small correction to the NB2071 photometry (by $-0.04$ mag), based on the median $K_s-$\,NB2071 colours of bright sources with $m_{\textrm{NB2071}}=$\,17--20 mag. At the position of the PKS\,1138, we estimate the Galactic extinction to be $E(B-V)=0.04$ based on the dust map of \cite{sch98}. This corresponds to $A_B=0.17$ mag, $A_{z'}=0.06$ mag, $A_J=0.03$ mag, $A_{K_s}=0.015$ mag, and $A_{\rm NB2071}=0.01$ mag, assuming the extinction law of \cite{car89}. 

\subsection{Spitzer MIPS 24$\mu$m data}

We use the {\it Spitzer} MIPS (\citealt{rie04}) 24$\mu$m data retrieved 
from the {\it Spitzer} Heritage Archive. The 24$\mu$m filter covers 
20.8--25.8$\mu$m corresponding to the rest-frame 6.6--8.2~$\mu$m 
for $z=2.16$ galaxies, which is dominated by 7.7$\mu$m PAH emission line. 
The retrieved data consist of three AORs (14888704; 14888960; 14889216) 
observed under the observing programme 20593, and the data covers 
a $\sim$5$'\times$\,5$'$ field around the radio galaxy.
We find that the post-basic calibration data (PBCD) images created 
at {\it Spitzer} Science Center show large-scale sky patterns. To remove 
the artefacts, we apply the self calibration on the basic calibration 
data (BCD) images following the recipe in the MIPS Data Handbook. 
We then remosaiced the images using the {\sc mopex} software (\citealt{mak05}) 
with 1.25$''$ sampling scale. Source detection and photometry is 
performed with {\sc SExtractor}. The limiting flux is measured in the 
same way as we did for optical and NIR data (with 12$''$ diameter aperture). 
The resultant 3-$\sigma$ limiting flux is 39$\mu$Jy. 
We detect 164 sources in total at $>$3-$\sigma$ significance  
within the 5$'\times$\,5$'$ MIPS field, among which we find 15 are 
associated with H$\alpha$ emitters. This matching is based on the 
nearest optical/NIR counterpart to each source, and we caution that, 
as a result of the poor PSF of 24$\mu$m data, these identifications 
may be uncertain for heavily blended sources in some crowded regions. 
When determining the 24-$\mu$m fluxes, we use 12$''$ aperture photometry 
($\times$\,2 of PSF size) with an aperture correction of 1.7\,$\times$ 
following the MIPS Data Handbook.

%%%%%%%%%%%%%%%%%%%%%%%%%%%%%%%%%%%%%%%%%%%%%%%%%%%%%%%%%%%%%%%%%%%%%%%%%%
% Analysis
%%%%%%%%%%%%%%%%%%%%%%%%%%%%%%%%%%%%%%%%%%%%%%%%%%%%%%%%%%%%%%%%%%%%%%%%%%
\section{Analysis}
\label{sec:analysis}

\subsection{Selection of H$\alpha$ emitters}

\begin{figure}
   \begin{center}
    \leavevmode
    \vspace{-2.2cm}
    \rotatebox{0}{\includegraphics[width=8.8cm,height=8.8cm]{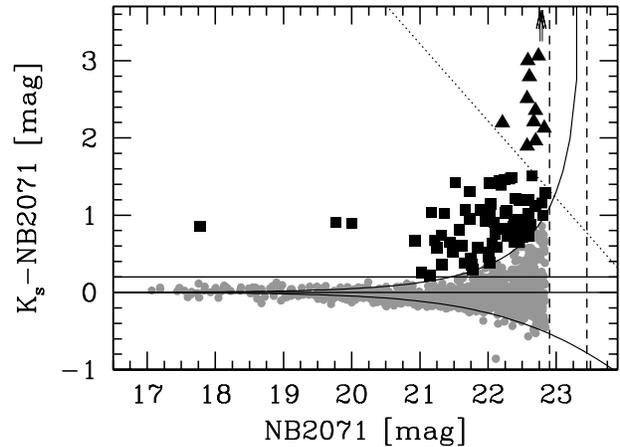}}
    \vspace{-0.7cm}
   \end{center} 
   \caption{The $K_s-$\,NB2071 versus NB2071 colour--magnitude diagram to define the NB2071 emitters. The vertical dashed-lines show 5-$\sigma$ and 3-$\sigma$ limiting magnitudes in NB2071, while the slanted dotted line shows the 2-$\sigma$ limiting magnitude in the $K_s$ band. The solid-line curves indicate $\pm 2.5\Sigma$ excess in $K_s-$\,NB2071 colours. Galaxies with $K_s-$\,NB2071\,$>0.2$ and $K_s-$\,NB2071\,$>2.5\Sigma$ are defined as the NB2071 excess sources (black symbols). The triangles show emitters with $K_s<2\sigma$, and two are above the boundary (shown as arrows). }
\label{fig:emitter_selection}
 \end{figure}

We select galaxies with a flux excess in the NB2071 filter compared to the $K_s$ filter. In Fig.~\ref{fig:emitter_selection}, we plot $K_s-$\,NB2071 colours of all the NB-detected sources against NB2071 magnitude. Considering the deviation of the plotted points around $K_s-$\,NB2071\,$=0$ and the photometric errors, we define NB2071 emitters as the galaxies which satisfy both (1) $K_s-$\,NB2071\,$>0.2$ and (2) $K_s-$\,NB2071\,$>2.5\Sigma$ (indicated by the solid-line curves in Fig.\ref{fig:emitter_selection}), where $\Sigma$ is the significance of the narrow-band excess (\citealt{bun95}). These criteria correspond to EW$_{\rm rest}\gsim20$\AA\ and $f$\,(H$\alpha$)$\gsim3\times10^{-17}$\,erg\,s$^{-1}$\,cm$^{-2}$ (which corresponds to a dust-free star formation rate of $\sim$\,10$M_{\odot}$\,yr$^{-1}$; see \S3.3). We find 83 sources satisfying these criteria  in total (Fig.~\ref{fig:emitter_selection}), among which 12 are fainter than 2-$\sigma$ limit of our $K_s$-band data. We do not use these $K_s$-undetected sources in the following analyses (except that we show their spatial distribution in \S4.1), because it is impossible to accurately measure their $J-K_s$ colours, stellar masses or star formation rates. We confirmed that omitting these faint sources does not affect our results. 

We checked that all nine spectroscopic members at $z\simeq 2.16$ 
with visible H$\alpha$ emission presented by \cite{kur04b} (except 
for one blended source in our NB image) are indeed selected as NB2071 
emitters in our analysis, confirming the reliability of our emitter 
selection procedure. In contrast, a majority of the spectroscopically
confirmed Ly$\alpha$ emitters (\citealt{pen00}) do not show detectable 
H$\alpha$ emission. This is not surprising because the nature of Ly$\alpha$ 
emitters and H$\alpha$ emitters may be different, and this is in fact 
the strong motivation to study distant galaxies using the H$\alpha$ line, 
which allows us to select analogs of local, star-forming galaxies. 
We also cross-checked with the photometric sample of H$\alpha$ emitter 
candidates shown in \cite{kur04a}. Out of the 39 H$\alpha$ emitter candidates 
listed in \cite{kur04a}\footnote{\cite{kur04a} listed 40 H$\alpha$ 
emitter candidates, while they suggest that one of their candidates 
(ID29) is a low-redshift interloper. We exclude this source from our analysis.}, 
20 have unique counterparts in our NB-selected catalogue, while the 
remaining 19 are too faint or blended in our Subaru data. We find that 
18 out of their 20 candidates in our catalogue (90\%) show a NB excess in 
our analyses. The missing two sources are the H$\alpha$ emitter candidates  
ID5 and ID154 in \cite{kur04a}. For ID154, we can still see an excess 
in our narrow-band data (and the source is indeed located near the 
selection boundary in Fig.~\ref{fig:emitter_selection}). 
On the other hand, the ID5 source shows no narrow-band excess 
($K_s-$NB$\sim$0) in our analysis. The reason for this is not clear, 
but it may be that the photometry for this source is less accurate 
in \cite{kur04a} because it is located near the edge of their VLT survey field.

%------------------------------------------------------------------------
%------------------------------------------------------------------------

%ZZZ

\subsection{Reliability of the selection}

\begin{figure}
   \begin{center}
    \leavevmode
    \vspace{-0.5cm}
    \rotatebox{0}{\includegraphics[width=8.7cm,height=8.7cm]{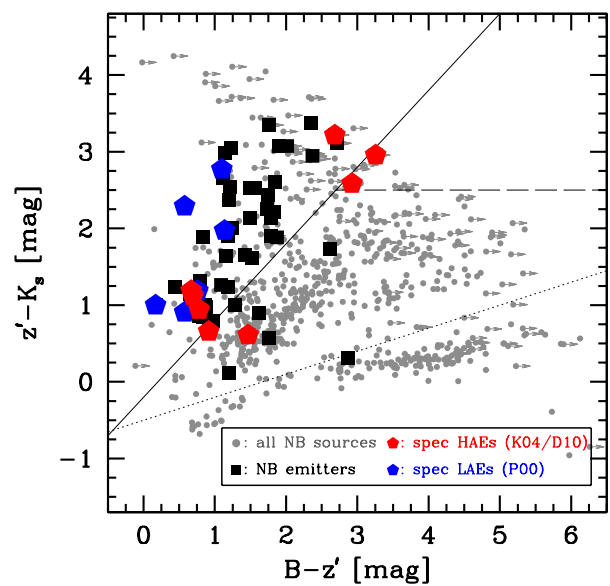}}
   \end{center} 
    \vspace{-0.6cm}
   \caption{The {\it BzK} diagram to check the colours of NB2071 emitters. The grey dots show all the NB-selected sources which are detected at both $z'$ and $K_s$, within the $B$-band data coverage. For $B$-undetected sources, we replaced their $B$-band magnitudes with  the 2-$\sigma$ limiting magnitude, and show their lower limits of $(B-z')$ colours with the right-ward arrows. The NB2071 emitters are shown with black squares. The red and blue pentagons are spectroscopically confirmed H$\alpha$ emitters (from Kurk et al.\ 2004b and Doherty et al.\ 2010) and Ly$\alpha$ emitters (from Pentericci et al.\ 2000), respectively. }
\label{fig:BzK}
 \end{figure}

The narrow-band excess sources selected above may contain a fraction of foreground or background contamination. The narrow-band technique guarantees an excess of emission at $\lambda = 2.07\mu$m in the selected NB emitters, but this could be caused by the \oiii\ line from emitters at $z\simeq 3.13$, \oii\ emitters at $z\simeq 4.55$, or the Pa$\alpha$ or Pa$\beta$ emitters at lower redshifts (e.g.\ \citealt{gea08}; \citealt{sob12}). To check the reliability of our H$\alpha$ emitter selection at $z=2$, we show the {\it BzK} diagram for these sources in Fig.~\ref{fig:BzK}. The {\it BzK}-selection technique (\citealt{dad04}) is designed to select distant galaxies over a relatively broad range in redshift ($1.4\lsim z \lsim 2.5$), but the major contamination concerned above would fall outside this range. Fig.~\ref{fig:BzK} clearly shows that a majority of NB2071 emitters  satisfy the {\it BzK} criteria, suggesting that they are really H$\alpha$ emitters at $z=2.16$. Also, most of the NB emitters are located in the top-left region of the {\it BzK} diagram (i.e.\ satisfying the {\it sBzK} criteria), supporting that they are star-forming population. Excluding the sources close to bright stars (for which $B$/$z'$ photometry is not possible), we find that 48 out of 56 (85\%) NB2071 emitters detected in both $z'$ and $K_s$ within the $B$-band data coverage satisfy the {\it BzK} selection criteria. Among the eight sources which fall outside the {\it BzK} criteria, four still satisfy the {\it BzK} criteria within their photometric errors. We also check the completeness of the {\it BzK} criteria using the spectroscopically confirmed members. We show in Fig.~\ref{fig:BzK} the {\it BzK} colours of 14 spectroscopic members of the PKS\,1138 structure reported in \cite{pen00}, \cite{kur04b}, and \cite{doh10} and detected in our catalogue. All the members  satisfy the {\it BzK} criteria (within the errors) as expected, except for  one X-ray detected galaxy at $z=2.157$. This should not be surprising as the strong active galactic nucleus (AGN) activity in this galaxy affects its broadband colours. 

The {\it BzK} selection may not be entirely suitable for excluding the high-redshift contamination (i.e.\ [O{\sc iii}] emitters at $z\sim 3$) which may be a concern. This would require deep spectroscopy for all of the sources to fully understand the contamination rates, but this is not available. We therefore check the photometric redshifts (photo-$z$) of our NB emitters derived by \cite{tan10b}. Due to the lack of multiband photometry in the southern half of our field, we can derive photometric redshifts for only 40 bright emitters (around half of our full H$\alpha$ emitter sample) located in the northern half of our survey. Although it is not easy to derive accurate photo-$z$'s for $z\gsim 2$ galaxies, particularly for star-forming galaxies with relatively flat spectral energy distributions (SEDs), our photometric redshift analysis shows that $>$\,70\% of emitters have $z_{\rm phot}=$\,1.8--2.4 (the majority of outliers show $z_{phot}<1.8$), supporting our claim that a majority of NB2071 emitters are H$\alpha$ emitters at $z=2.16$. The contamination rate may of course be higher in the southern part of our survey field. However, a similar depth narrow-band study of $z=2.23$  H$\alpha$ emitters by \cite{gea08} estimate that the high-redshift contamination (i.e.\ [O{\sc iii}] emitters at $z\sim 3$) into the NB emitters (selected at $\sim$\,2$\mu$m) is negligibly small, by applying the Lyman Break Galaxy (LBG) technique (they find only one source satisfying the LBG selection from 180 emitters in their whole  sample). Based on this estimate, we expect that the high-redshift contamination into our sample would be negligibly small. In summary, we estimate the total contamination rate would be $\lsim$\,10\%, even if we consider all non-{\it BzK} galaxies as contamination. In this paper, we assume all the 83 NB emitters are H$\alpha$ emitters at $z=2.16$, but the results do not change if we apply either the {\it BzK} or photo-$z$ selections.  

%------------------------------------------------------------------------
%------------------------------------------------------------------------

\subsection{AGN contamination}

Recent studies showed that the fraction of AGN increases towards high-redshift clusters (e.g. \citealt{eas07}; \citealt{mar09}; \citealt{tom11}). Although we expect that the majority of our H$\alpha$-selected sources are star-forming galaxies, a fraction of H$\alpha$ emitters might be powered by AGNs, and so we here check the possible contribution of AGNs into our sample. Firstly, we match our sample with the {\it Chandra} X-ray catalogue (down to a 0.5--2 keV flux of $\sim$\,10$^{-15}$ erg\,s$^{-1}$\,cm$^{-2}$) presented by \cite{pen02}, and find that five of our H$\alpha$ emitters are detected in the soft X-ray band: the source IDs of \#3, \#5, \#6, \#7 (=radio galaxy), and \#17 in the nomenclature of \cite{pen02}. We find that all of them are known proto-cluster members from the subsequent NIR/optical spectroscopy (\citealt{kur04b}; \citealt{cro05}), and so our survey confirms H$\alpha$ emission from all of them, while we do not find any additional candidate X-ray AGNs. 

Secondly, we check the SEDs of the H$\alpha$ emitters, using the best-fit SEDs derived from the updated code of \cite{tan10b}. The SED fitting was performed for 40 H$\alpha$ emitters in the PKS\,1138-C field where multi-band data are available. Although we are aware of significant uncertainties in our SED fitting, we find that most of our H$\alpha$ emitters are fit by star-forming galaxy templates, and that passive SEDs are a minority. We also check the dust extinction ($A_V$) estimated from the SED fitting. An accurate estimate of dust extinction from SED fitting is extremely difficult, due to the degeneracies of age, metallicity and dust extinction, but we find a weak trend that those H$\alpha$ emitters with red $(J-K_s)$ colours tend to display higher extinction ($A_V\gsim 1$ mag), compared to blue galaxies ($A_V\lsim 1$ mag). This would also suggest that the red H$\alpha$ emitters are likely to be star forming, rather than dust-free, passive galaxies. Therefore, in this paper, we assume all the H$\alpha$ emitters (including five X-ray detected sources) are star forming, although we cannot completely rule out the presence of AGNs without deep spectroscopy for all of them.

\subsection{Stellar mass}

The transformation from rest-frame luminosity to stellar mass has the least dispersion (due to different star formation histories and reddening) in the NIR bands. Unfortunately, the {\it Spitzer} IRAC (restframe NIR at $z=2.16$) coverage of our field is incomplete and so we choose to use the more complete  $K_s$-band imaging to estimate stellar masses. We confirm that the two techniques agree by comparing the stellar masses derived with the simple $K_s$-band method described below, with those derived from the SED fitting including {\it Spitzer} IRAC photometry by \cite{tan10b} (using the \citealt{bc03} stellar population synthesis model). We find excellent agreement, at least for the brightest sources where the SED fitting is possible. Hence, as the SED fitting can only be performed for  the brighter sources at the northern half of our survey (those with IRAC coverage), we use the $K_s$-band estimates for all our sample in this paper to keep consistency throughout the field. 

The stellar mass estimates from the $K_s$-band use the observed galaxy luminosity and take into account the variation in mass to light ratio ($M/L$) of the galaxies, dependent upon their star formation history (or equivalently their observed colours), by exploiting  the following equation:
\begin{equation}
\log (M_*/10^{11}M_{\odot}) = -0.4(K_s-22.24) + \Delta\log M,
\end{equation}
which  is derived by simple scaling of the model galaxies with $M_*=10^{11}M_{\odot}$ (with $K_s=22.24$) from \cite{kod99}, assuming a formation redshift of $z_f=5$. The colour dependence of mass-to-light ratio ($M/L$) is included in the last term, $\Delta\log M= 0.03$--$1.5\times \exp[-1.11\times (z'-K_s)]$, which is derived by fitting the model galaxies with a range of star-formation histories (\citealt{kod99}). We use $(z'-K_s)$ colour which neatly straddles the redshifted 4000\AA\ break providing a sensitive tracer of the typical age of the stellar populations in these galaxies. Unfortunately we find that there are two H$\alpha$ emitters near a bright star for which accurate optical photometry is not possible. For these two sources we instead use the $(J-K_s)$ colour to predict the $M/L$ using the same model with $\Delta\log M = 0.14$--$0.9\times \exp(-1.23\times (J-K_s))$. We verified that this method provides consistent results with those derived using $(z'-K_s)$ colour. We also rescale the stellar mass using the \cite{sal55} initial mass function (IMF) for consistency with the derivation of SFRs as described below. Finally, we correct for the stellar mass accounting
for the contribution of H$\alpha$ emission line flux to the broad-band photometry, using our measured NB fluxes.

Applying different models or IMFs would slightly change the absolute measurements of $M_*$, but the trend we show in this paper does not change if we apply different models/IMFs. Indeed, we verified that our conclusions do not alter if we apply an empirical conversion from $K$-band photometry to $M_*$ (with $M/L$ correction depending on $J-K_s$ colour) using the SED fitting results for $z=2.2$ H$\alpha$ emitters from HiZELS (D.\ Sobral; private communication), which assumes the updated version of \cite{bc03} model and \cite{cha03} IMFs. 

%------------------------------------------------------------------
  \begin{figure*}
   \vspace{-0.3cm}
   \begin{center}
    \leavevmode
    \epsfxsize 0.48\hsize
    \epsfbox{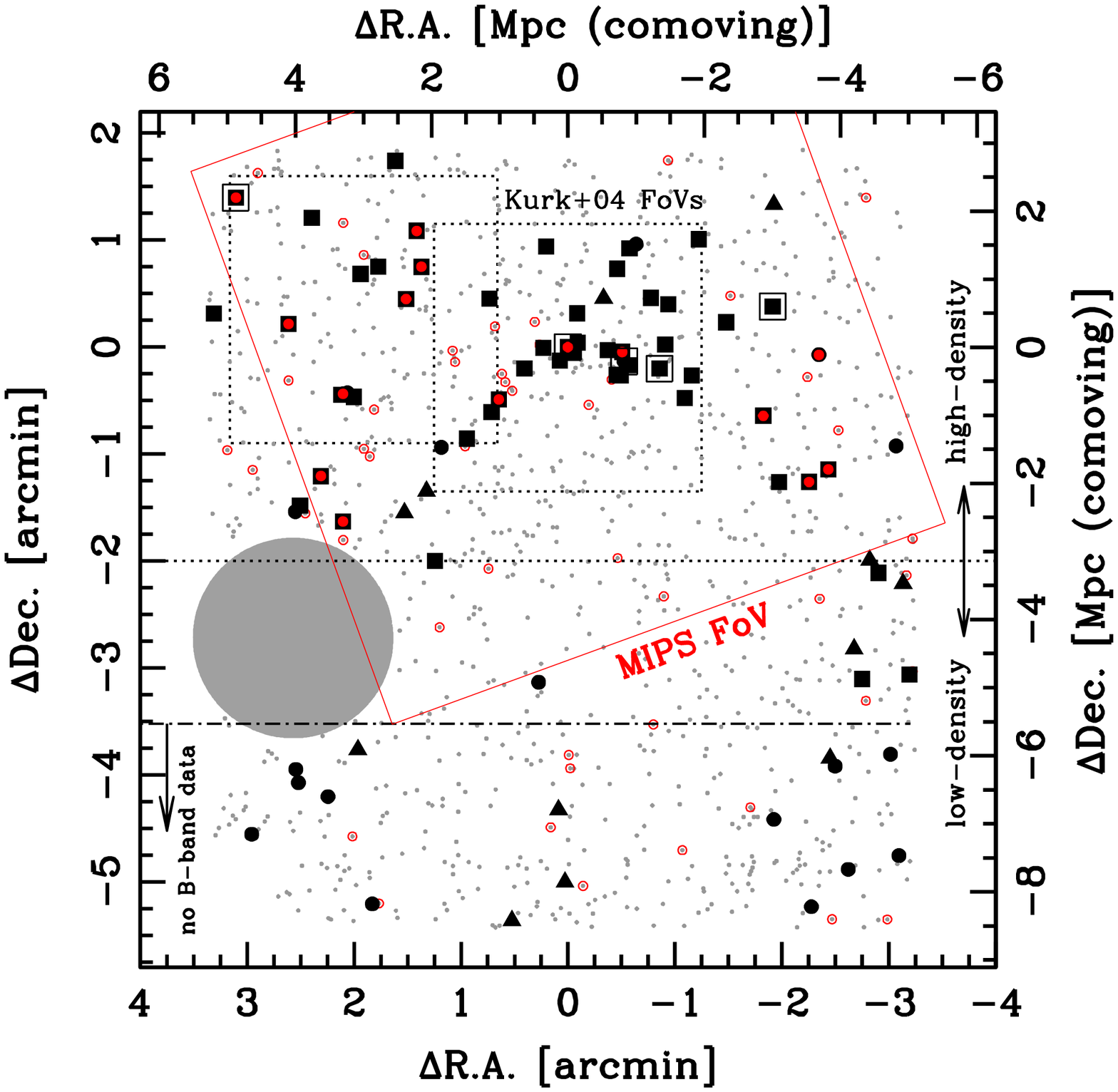}
    \hspace{1mm}
    \epsfxsize 0.48\hsize
    \epsfbox{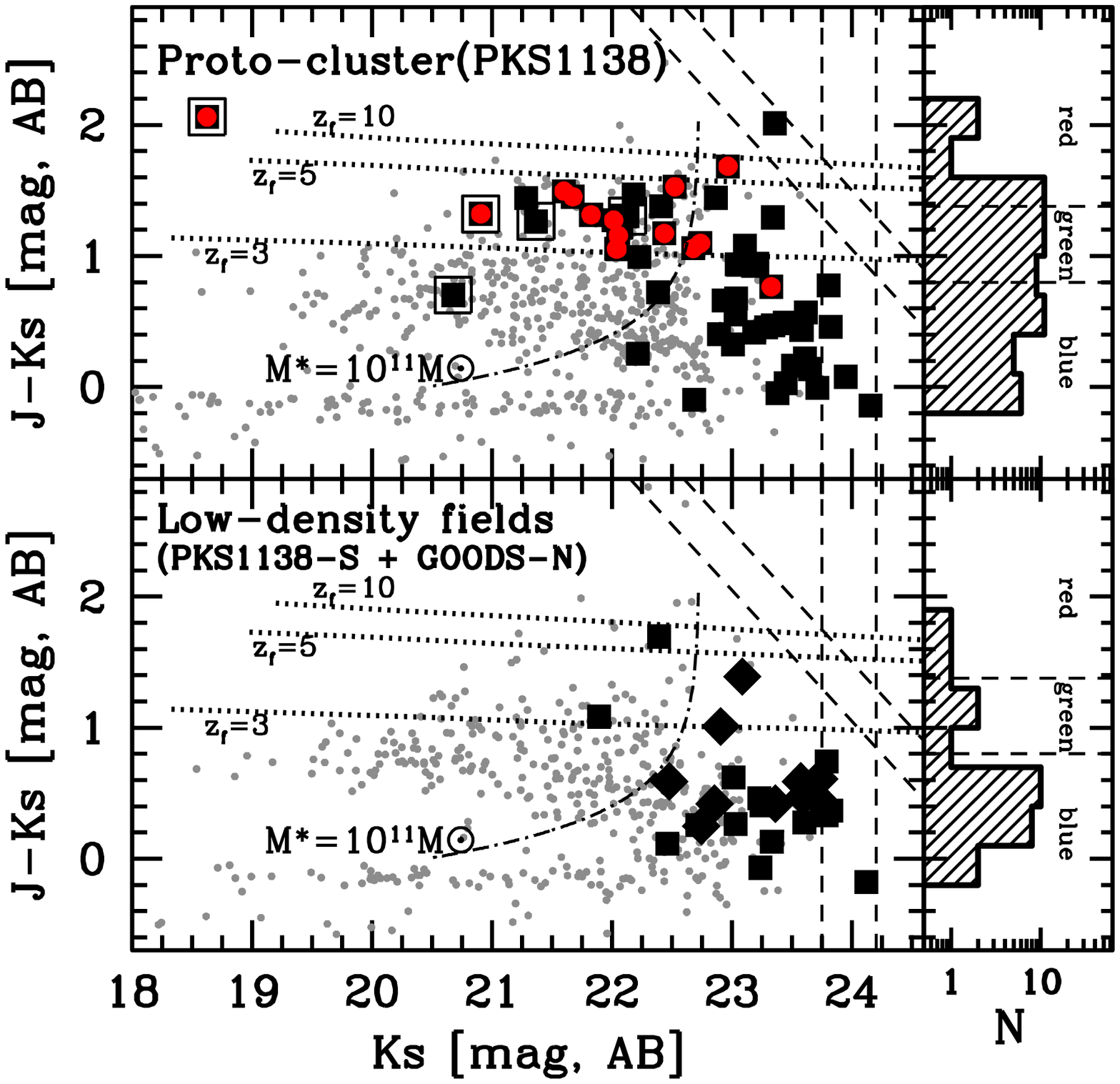}
   \end{center} 
  \vspace{-3mm}
   \caption{({\it Left}): The 2-D map of the H$\alpha$ emitters around PKS\,1138. The black squares show H$\alpha$ emitters which satisfy {\it BzK} selection, while black circles show H$\alpha$ emitters which do not satisfy the {\it BzK} selection. The triangles show the $K_s$-undetected emitters. The red filled circles and black open squares show MIPS and X-ray detected emitters, respectively. Grey points show all the NB2071-selected sources, and DRGs are marked with red open circles. Note that the $B$-band data is only available at $\Delta$Dec.$\ge$$-3.5$, so that we do not have {\it BzK} information below this line. The coordinates are given relative to the central radio galaxy. The large grey circle shows the masked region due to some bright sources. The dotted line at $\Delta$Dec.$=$$-2.0$ shows the dividing line of high- and low-density environment used in this paper. ({\it Right}): The $(J-K_s)$ vs $K_s$ colour--magnitude diagram for the high-density proto-cluster and the low-density fields. The grey dots indicate all the NB-selected sources, and the black squares are H$\alpha$ emitters (red filled circles for MIPS-detected emitters). The diamonds in the bottom panel are the $z=2.2$ H$\alpha$ emitters in the GOODS-N field. The vertical and slanted dashed lines show 3-$\sigma$ and 2-$\sigma$ limiting magnitudes of $K_s$-band and $J$-band data, respectively. The locations of the red sequence in the case of $z_f = 3$, 5, 10 modeled by \citet{kod97} are shown as the dotted lines, and the dash--dotted line shows the iso-stellar-mass line (corresponding to $10^{11}M_{\odot}$) based on the model of \citet{kod99}. The colour distribution of emitters within each environment is shown as the histogram, where we also show our colour definitions. }
\label{fig:map_all}
 \end{figure*}
%-----------------------------------------------------------------

\subsection{Star formation rate}

Our narrow-band H$\alpha$ imaging allows us to measure the emission line strengths and hence SFRs of all the sources. We derive the SFRs by estimating the line flux ($F_{\rm{H\alpha+[NII]}}$), continuum flux density ($f_c$), and the rest-frame equivalent width (EW$_{\rm rest}$) with the following equations\footnote{During this process, we find a source located near the central radio galaxy which is estimated to have an unrealistic negative continuum level. It appears that the {\sc mag\_auto} photometry is failing for this source because of the extended light from the central radio galaxy. We therefore decided to exclude this source from the following analyses where stellar mass or star formation rate is needed.}: 
\begin{equation}
F_{\rm{H\alpha + [NII]}}=\Delta_{\rm{NB}}\frac{ f_{\rm{NB}} -
 f_{K_s} }{1-\Delta_{\rm{NB}}/\Delta_{K_s}}
\end{equation}
\begin{equation}
f_c = \frac{f_{K_s} - f_{\rm{NB}}(\Delta_{\rm{NB}}/\Delta_{K_s})}{1-\Delta_{\rm{NB}}/\Delta_{K_s}}
\end{equation}
\begin{equation}
\textrm{EW}_{\rm rest}({\rm{H\alpha + [NII]}}) = (1+z)^{-1} \frac{F_{\rm{H\alpha}
 + [NII]}}{f_c} 
\end{equation}
where $\Delta_{K_s}$ ($=0.31\mu$m) and $\Delta_{\rm{NB}}$ ($=0.027\mu$m) are the FWHMs of the $K_s$ and NB2071 filters, and $f_{K_s}$ and $f_{\rm{NB}}$ are the flux densities at $K_s$-band and at NB2071, respectively. We then transform $F_{\rm{H\alpha +[NII]}}$ to $L_{\rm{H\alpha +[NII]}}$. Finally, we compute the H$\alpha$-based star formation rates (SFR$_{\rm{H}\alpha}$) using the relation of \cite{ken98} assuming the \cite{sal55} IMF; SFR$_{\rm{H}\alpha}$(M$_{\odot}$\,yr$^{-1}$)\,$=7.9\times 10^{-42} L_{\rm{H\alpha}}$ (erg\,s$^{-1}$). We here adopt an empirical correction of the [N{\sc ii}] line contribution to the measured line fluxes using the relation between [N{\sc ii}]/H$\alpha$ and the rest-frame EW$_{\rm{H}\alpha+\rm{[NII]}}$ presented in \cite{sob12}, as well as the stellar mass-dependent dust extinction correction shown in \cite{gar10b} after taking the IMF difference into account (\citealt{gar10b} assumed \citealt{kro01} IMF, and so we scaled their stellar mass by $+$0.24\,dex). We checked that most of our results are unchanged even if we applied the conventional constant 1-mag correction or a SFR-dependent extinction correction (e.g.\ \citealt{gar10a}), but some possible effects on our results will be discussed in \S4.4.4.

For the MIPS-detected emitters, we also estimate the IR-derived star formation rates (SFR$_{\rm IR}$) using the 24$\mu$m photometry. We first estimate the total infrared luminosity ($L_{\rm IR}$) using the model SEDs of starburst galaxies (\citealt{lag03}), and then transform to SFR$_{\rm IR}$ using the \cite{ken98} relation; SFR$_{\rm IR}$~(M$_{\odot}$\,yr$^{-1}$) $=4.5\times10^{-44}L_{\rm IR}$\,(erg\,s$^{-1}$). We find that the MIPS-detected emitters have SFR$_{\rm IR}\gsim$\,100\,M$_{\odot}$\,yr$^{-1}$, which are typically factor $\sim$\,2--3 higher than the H$\alpha$-derived SFR. Therefore it is possible that the $M_*$-dependent extinction correction applied above may be still underestimating the extinction correction in the massive galaxies (note that the MIPS-detected emitters are most massive emitters with $M_*\gsim 10^{11}M_{\odot}$; see \S4.2). In this paper, we use the H$\alpha$-derived SFRs, due to the limited number of the MIPS-detected emitters and the restricted MIPS data coverage, and due to the large uncertainty in deriving $L_{\rm IR}$ from a single rest-frame $\sim$8$\mu$m photometric point (e.g.\ \citealt{nor10}).

%%%%%%%%%%%%%%%%%%%%%%%%%%%%%%%%%%%%%%%%%%%%%%%%%%%%%%%%%%%%%%%%%%%%%%%%%%
% 4. RESULTS & DISCUSSIONS
%%%%%%%%%%%%%%%%%%%%%%%%%%%%%%%%%%%%%%%%%%%%%%%%%%%%%%%%%%%%%%%%%%%%%%%%%%
\section{Results and Discussions}
\label{sec:lss}

\subsection{Large-scale structure around PKS\,1138}

We show in Fig.~\ref{fig:map_all} the spatial distribution of the H$\alpha$ emitters as well as the MIPS and X-ray detected emitters. It is clear that the H$\alpha$ emitter candidates are highly concentrated around the  radio galaxy, which is consistent with the previous H$\alpha$ study of this field by \cite{kur04a}. In contrast, the number density of H$\alpha$ emitters in the southern part of the field is significantly lower. To quantify this we estimate density of H$\alpha$ emitters in the proto-cluster core region (within 40 arcsec from the radio galaxy) as $\sim$\,10.0 arcmin$^{-1}$, which is $\sim$\,10 times higher than the average number density calculated in the southern half of our survey.  Thus, our findings confirm the many previous studies which have claimed the presence of an over-density of galaxies in the PKS\,1138 field from various techniques (e.g.\ \citealt{pen00}; \citealt{kur04a}; \citealt{cro05}; \citealt{kod07}; \citealt{tan10b}).

We also see from Fig.~\ref{fig:map_all} that the PKS\,1138 proto-cluster is actually embedded in a very large-scale filament, running from north-east to south-west  through the cluster centre. The north-east filament has already been reported by \cite{kur04a}, but our new survey has revealed an even larger-scale filament ($\gsim$\,10 Mpc in comoving scale) traced by H$\alpha$ emitters, which also possibly extends off to the south east (although, unfortunately, masking due to bright objects is hiding some of the structure). Such large-scale structures in the nearby Universe are generally traced by passive galaxies, but the situation is not necessarily the same in the distant Univese, where star-forming population are also highly clustered in the proto-cluster core (see consistent results from $1.5\lsim z \lsim 2.5$ cluster studies by \citealt{hay10}; \citealt{hil10}; \citealt{tra10}). Nevertheless, the structure around the PKS\,1138 is also suggested from an analysis of distant red galaxies (DRGs) by \cite{kod07}, although the structure traced by H$\alpha$ emitters presented in this study appears to be much more prominent. This may be because the DRG selection picks up distant galaxies over a wide range in redshift (e.g.\ \citealt{fra03}), diluting the contrast of any structures. We check the spatial distribution of DRGs throughout the field (Fig.~\ref{fig:map_all}) and confirm that the spatial distribution of DRGs is qualitatively consistent with that of H$\alpha$ emitters, but again, the structure looks less prominent. 

\subsection{Red star-forming galaxies and their environment}

\subsubsection{Colour--magnitude diagram}

We show in Fig.~\ref{fig:map_all} the colour--magnitude diagrams for high- and low-density environment separately. Considering the complex structure around PKS\,1138, we simply define the upper and lower half of our survey field as high- and low-density region, respectively (see Fig.~\ref{fig:map_all}). Because the number of H$\alpha$ emitters in the low-density environment is much smaller than that in the high-density region, we included in our analysis the $z=2.2$ H$\alpha$ emitter sample from the blank field survey in the GOODS-North field with Subaru by \cite{tad11}. Although the NB filter used in \cite{tad11} is different (NB209 filter with $\lambda_c=2.09\mu$m), the H$\alpha$ emitter selection technique and all the broad-band filters used in that study are exactly the same as those used in this study. After applying the same flux and EW cut, we add these sources in our field galaxy sample. 

%------------------------------------------------------------------
  \begin{figure*}
   \vspace{-0.3cm}
   \begin{center}
    \leavevmode
    \epsfxsize 0.49\hsize
    \epsfbox{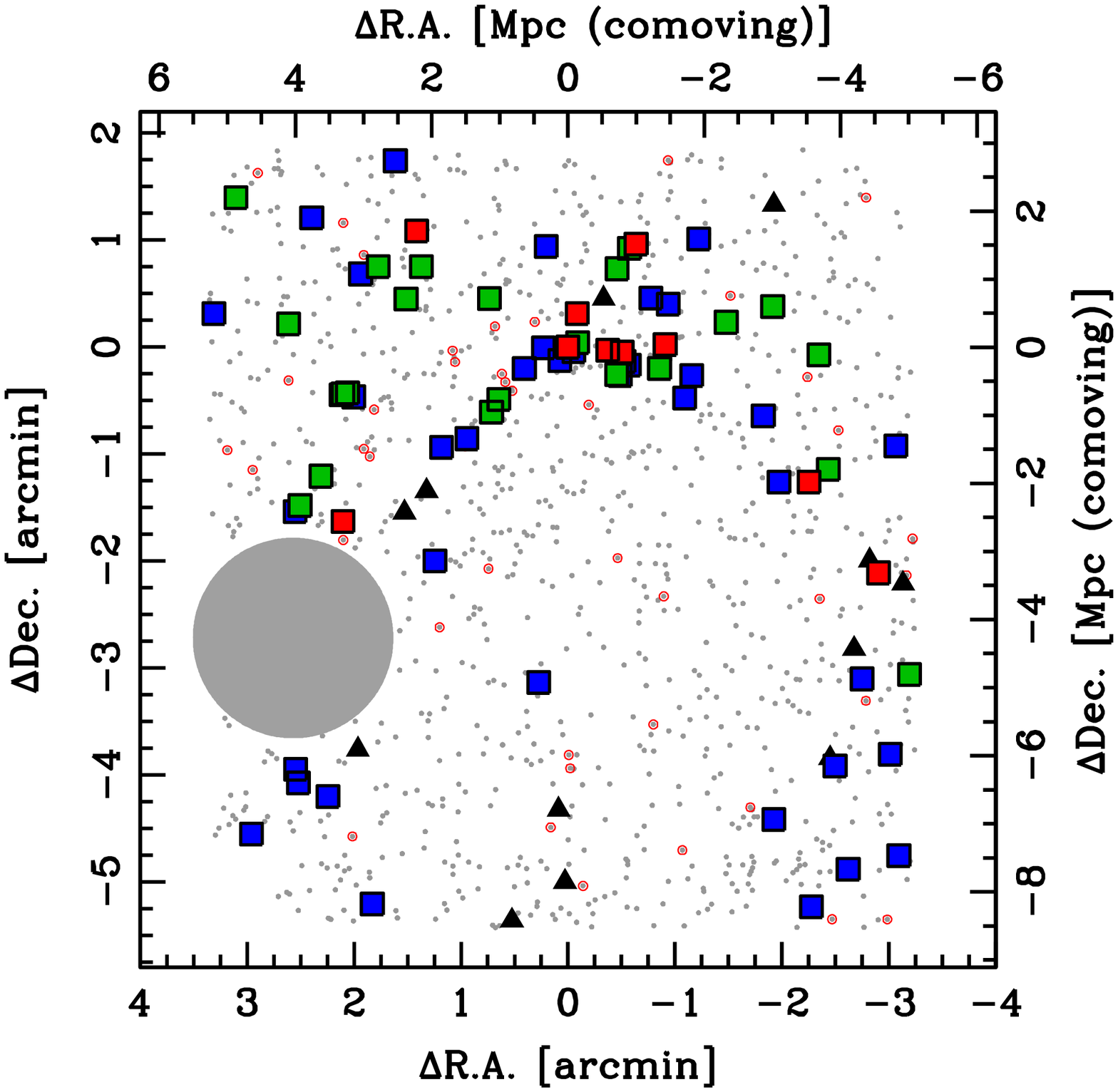}
    \epsfxsize 0.49\hsize
    \epsfbox{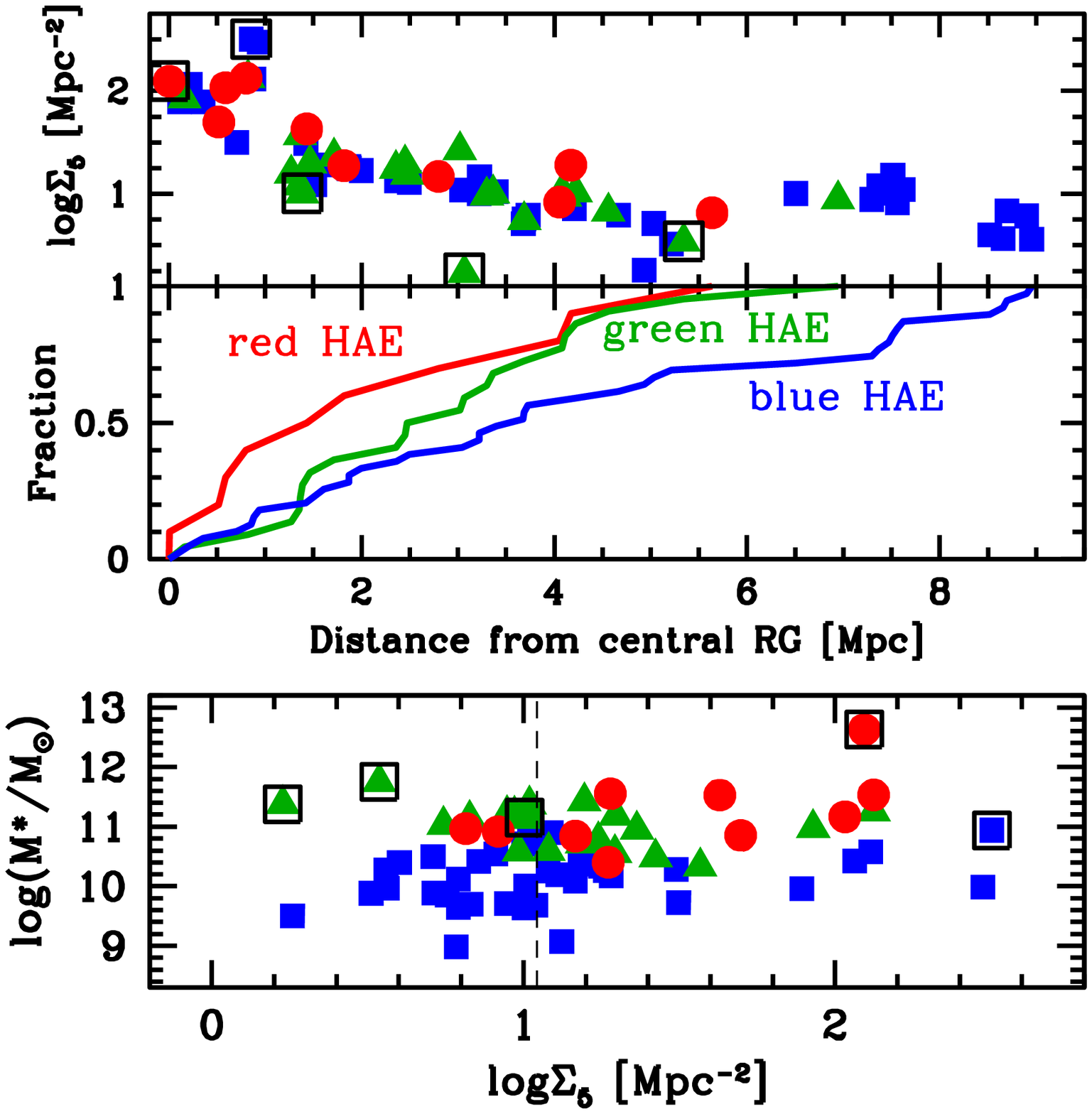}
   \end{center} 
  \vspace{-3mm}
   \caption{ ({\it Left}): The same plot as Fig.~4, but using different colour symbols based on their $(J-K_s)$ colour. The red, green and blue squares indicate H$\alpha$ emitters with $(J-K_s)>1.38$ (DRGs), $0.8<(J-K_s)<1.38$ and $(J-K_s)<0.8$, respectively (see text). The black triangles are the $K_s$-undetected emitters. 
({\it Top-right}): The cumulative fraction and the local number densities (calculated with all emitters) of red, green, and blue emitters as a function of distance from the central radio galaxy. The different colour symbols indicate different $(J-K_s)$ colours, and the open squares show the X-ray detected emitters. ({\it Bottom-right}): The stellar mass of emitters as a function of the local density. The meanings of the symbols are the same as those in the top panel. The vertical dashed line shows the median value of $\log_{10} \Sigma_5=1.04$.}
\label{fig:map_color}
 \end{figure*}
%-----------------------------------------------------------------

In Fig.~\ref{fig:map_all}, we find a large number of ``red H$\alpha$ emitters'' with $(J-K_s)\gsim 1$. In some extreme cases, they satisfy the DRG selection criteria with $(J-K_s) > 1.38$ (equivalently, $(J-K_s)_{\rm{(Vega)}}>2.3$). We confirmed that the two red sources with H$\alpha$ emission shown in the spectroscopic study by \cite{doh10} are both selected as red emitters in this study. Importantly, we find that such red H$\alpha$ emitters are predominantly found in the proto-cluster environment, while they are very rare in the low-density environment. A Kormogorov-Smirnov test (KS test) shows that the probability that the colour distribution of the high-density and low-density samples are drawn from the same parent population is only 0.1\%. We note that this result may partly be produced by the presence of X-ray detected AGNs which also tend to be red and massive (see Fig.~\ref{fig:map_all}), but the result does not change even if we exclude all the X-ray sources.

We also find that the red H$\alpha$ emitters tend to have systematically brighter $K_s$-band luminosities compared to blue emitters. We show in Fig.~\ref{fig:map_all} the iso-stellar-mass line corresponding to $M_*=10^{11}M_{\odot}$ based on the same model as that used when deriving $M_*$ (see \S3.4). It is notable that the H$\alpha$ emitters with red $(J-K_s)$ colours tend to have very high stellar masses with $M_*\gsim 10^{11}M_{\odot}$. These massive star-forming population are absent from lower-redshift clusters, and therefore we propose that they are the  progenitors of present-day passive cluster galaxies (see more detailed discussion in \S4.4). Our result suggests that the cluster galaxies had already formed a large part of their stellar mass by $z\sim 2$, but they are still in a vigorously star-forming phase. There is no strict definition of ``green'' galaxies, but we here distinguish relatively redder emitters with $(J-K_s)\sim1$ (which tend to be brighter in $K_s$-band) from the very blue emitters with $(J-K_s)\lsim 0.5$ (which tend to be fainter in $K_s$-band), as recognized on the colour--magnitude diagram. Hence, in the remainder of this paper, we define the red, green and blue emitters as those having $(J-K_s)>1.38$ (i.e.\ DRG), $0.8<(J-K_s)<1.38$ and $(J-K_s)<0.8$, respectively. 

We also show in the colour--magnitude diagram the MIPS-detected H$\alpha$ emitters (red circles in the top panel). Although the MIPS data only covers the higher-density region, it is notable that most of the MIPS-detected emitters are red/green massive sources. Within the MIPS data coverage, we find that 14 out of 29 (48\%) are detected at 24$\mu$m for red/green emitters, while only 1 out of 24 (4\%) is detected for blue emitter. This trend is of course related to the difference in the stellar mass distribution between red and blue sample (see also \S4.4), but this result suggests that the red emitters are likely dusty sources rather than passive galaxies. Their MIR-derived SFR is estimated to be SFR$_{\rm IR}\gsim 100$\,M$_{\odot}$\,yr$^{-1}$ (and with the dust extinction estimated to be $A_{\rm H\alpha}\sim 2$--3 mag), suggesting that these ULIRG-class active galaxies are a common population in proto-cluster environment at $z\sim 2$ (see also \citealt{tani11}). We note that the red H$\alpha$ sources without MIR detection also have strong H$\alpha$ emission with SFR$_{\rm H\alpha}\gsim$\,50$M_{\odot}$\,yr$^{-1}$, and so we expect that they are also strong starbursts rather than passive sources. Recently, \cite{may12} analysed the {\it Spitzer} MIPS data of a number of high-redshift radio galaxy fields including the PKS\,1138 field. That study found an overdensity of MIPS 24$\mu$m sources in the PKS\,1138 field at 4.3-$\sigma$ significance, but it has been impossible to study the MIR detection from individual cluster galaxies because of the difficulty of determining their membership. We note that the clean H$\alpha$ emitter sample presented here has allowed us to confirm the presence of MIR-bright sources in the PKS\,1138 proto-cluster. 

%------------------------------------------------------------------------
 \begin{figure*}
   \begin{center}
    \leavevmode
    \vspace{-1.0cm}
    \rotatebox{90}{\includegraphics[width=12.8cm,height=18cm]{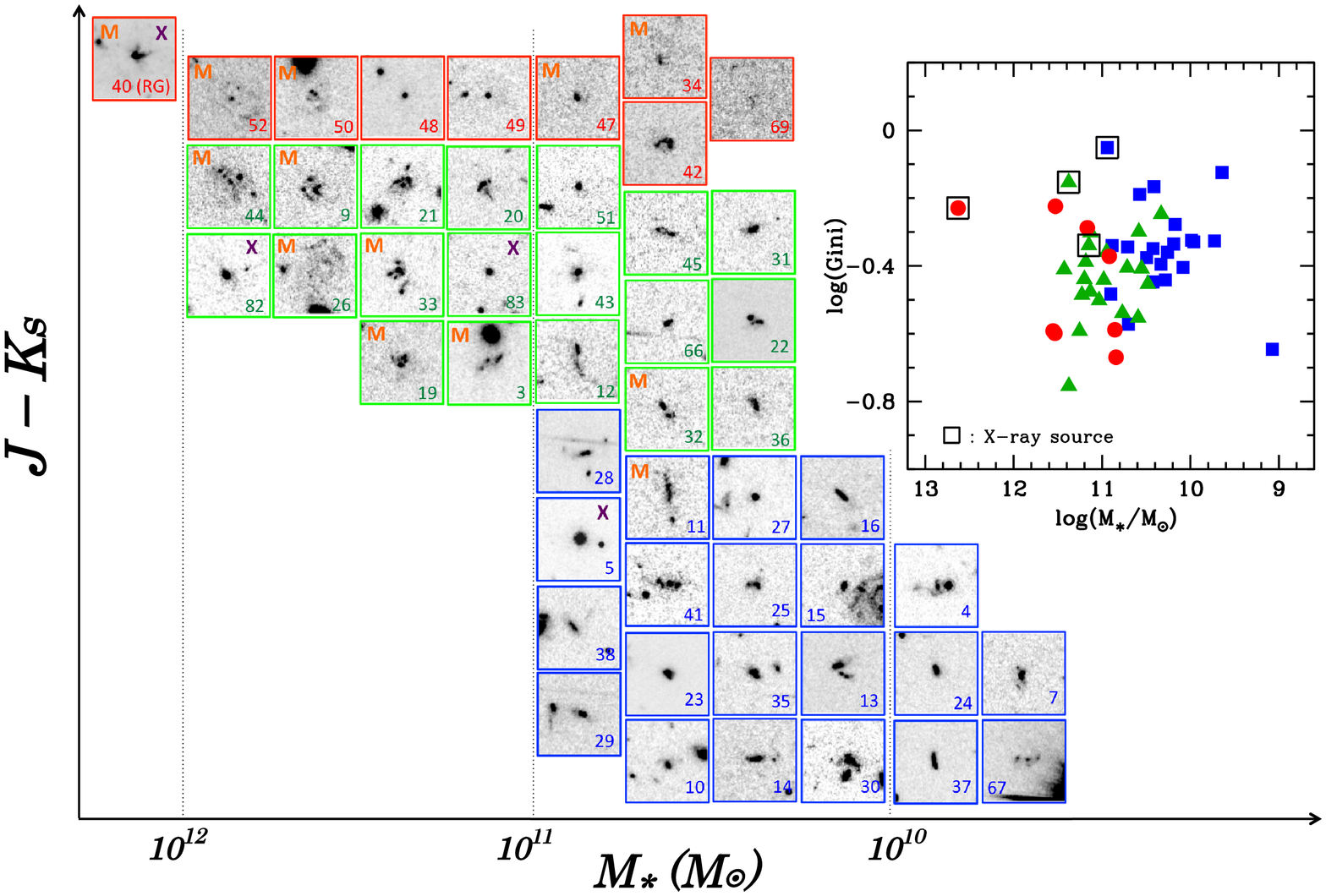}}
    \vspace{-1.8cm}
   \end{center} 
   \caption{ The {\it HST} ACS $I_{814}$-band snapshots (4$''\times$4$''$ for each) of the H$\alpha$ emitters in the PKS\,1138 field, plotted on the colour versus stellar mass plane (the IDs are given based on our H$\alpha$ emitter catalogue). Note that the locations of each source in this plot does not exactly reflect their stellar masses or colours in order to avoid overlap with each other. The colours of border lines of each snapshot indicate red, green, and blue emitters, as we used throughout the paper. The "X" and "M" denote the X-ray and MIPS detected sources, respectively. Note that we find five X-ray detected H$\alpha$ emitters, but one of them is outside of the {\it HST} data coverage. We also plot the Gini coefficients ({\it G}) of these galaxies as a function of stellar mass (the top-right panel). The red, green and blue symbols indicate their $(J-K_s)$ colours, and the open squares are for X-ray detected emitters. Note that we do not calculate {\it G} for ID69 which is too faint in the {\it HST} image. }
\label{fig:HST_snapshot}
 \end{figure*}
%------------------------------------------------------------------------

\subsubsection{Red star forming galaxies: a proto-cluster phenomenon?}

In Fig.~\ref{fig:map_color}, we show the spatial distribution of the red, green, and blue emitters defined above. This plot demonstrates our important finding that the redder sources are strongly clustered in the proto-cluster field. We here apply a more quantitative measurement of the environment to show the environment of the red massive H$\alpha$ emitters at $z=2$. At first, we plot in Fig.~\ref{fig:map_color} (top-right panel) the radial distribution of red, green, and blue H$\alpha$ emitters from the central radio galaxy. Using a Kolmogorov-Smirnov (KS) test, we find that the probability that red/green H$\alpha$ emitters and blue H$\alpha$ emitters are drawn from the same parent population is 4.6\%. We also find that 5 out of 10 red H$\alpha$ emitters (50\%) are clustered within 1 arcmin from the radio galaxy, while only 9 out of 39 blue H$\alpha$ emitters (23\%) are located in the same region, supporting an excess of red emitters in the proto-cluster environment. Next we calculate the local number density of H$\alpha$ emitters ($\Sigma_5$) and we plot them as a function of the distance from the radio galaxy (middle-right panel in Fig.~\ref{fig:map_color}). With this plot, we can conclude the over-density of H$\alpha$ emitters around PKS\,1138 (by a factor of $\sim$\,10 compared to its surrounding field; see also \S4.1). We also show in Fig.~\ref{fig:map_color} (bottom-right) a plot of the stellar mass of H$\alpha$ emitters as a function of the local density. Although the trend seems to be less prominent, we can still see a hint that redder H$\alpha$ emitters prefer the higher-density environment. In fact, by splitting the sample into high- and low-density bins at the median value of the $\Sigma_5$ (shown as the dashed line in Fig.~\ref{fig:map_color}), we find that 8 out of 10 red emitters (80\%) are located in the higher-density bin, 
supporting an excess of red, massive emitters in the proto-cluster environment. Note that as this density measurement is made using only the H$\alpha$ emitter sample (we do not know the location of passive galaxies in the proto-cluster),  we have decided to retain the more simple definition of environment for the remainder of this study (i.e.\ the high-density environment is defined as the northern half of our survey). The results shown here do not change if we use the density-based definition. 

Our finding of an over-density of red H$\alpha$ sources near the proto-cluster core is qualitatively consistent with our recent study of a proposed proto-cluster field at $z=2.53$ by \cite{hay12}, who also found a concentration of red H$\alpha$ emitters in high-density clumps. However, it would be interesting to note that the very massive ($M_*\gsim 10^{11}M_{\odot}$) red emitters are not a common population in that study. Such very massive red star-forming galaxies are also very rare in another proto-cluster field 4C+10.48 at $z=2.35$ studied by \cite{hat11}. The colour-magnitude diagram shown in \cite{hat11} also suggests that red emitters themselves are very rare in the 4C+10.48 field, and in fact they suggested that there is no strong environmental variation in the colour distribution of H$\alpha$ emitters at $z\sim 2$. These different results between the proto-cluster fields would indicate that the cluster-to-cluster variation is significant in the early Universe. Interpretation of this difference is difficult at this point, but it may be  that the properties of galaxies are related to the richness of clusters or surrounding structure in the sense that more mature clusters tend to harbour redder and more massive galaxies inside them. Perhaps, the PKS\,1138 field may be an unusually rich field at $z\sim 2$,  as suggested by the huge surrounding structure, which may have then collapsed early on. In any case, it is clearly important to study larger sample of $z\gsim 2$ proto-cluster galaxies in the future to obtain firmer conclusions on environmental variations of galaxy properties at such high redshifts.

%------------------------------------------------------------------------
 \begin{figure*}
   \vspace{-7mm}
   \begin{center}
    \leavevmode
    \epsfxsize 0.48\hsize
    \epsfbox{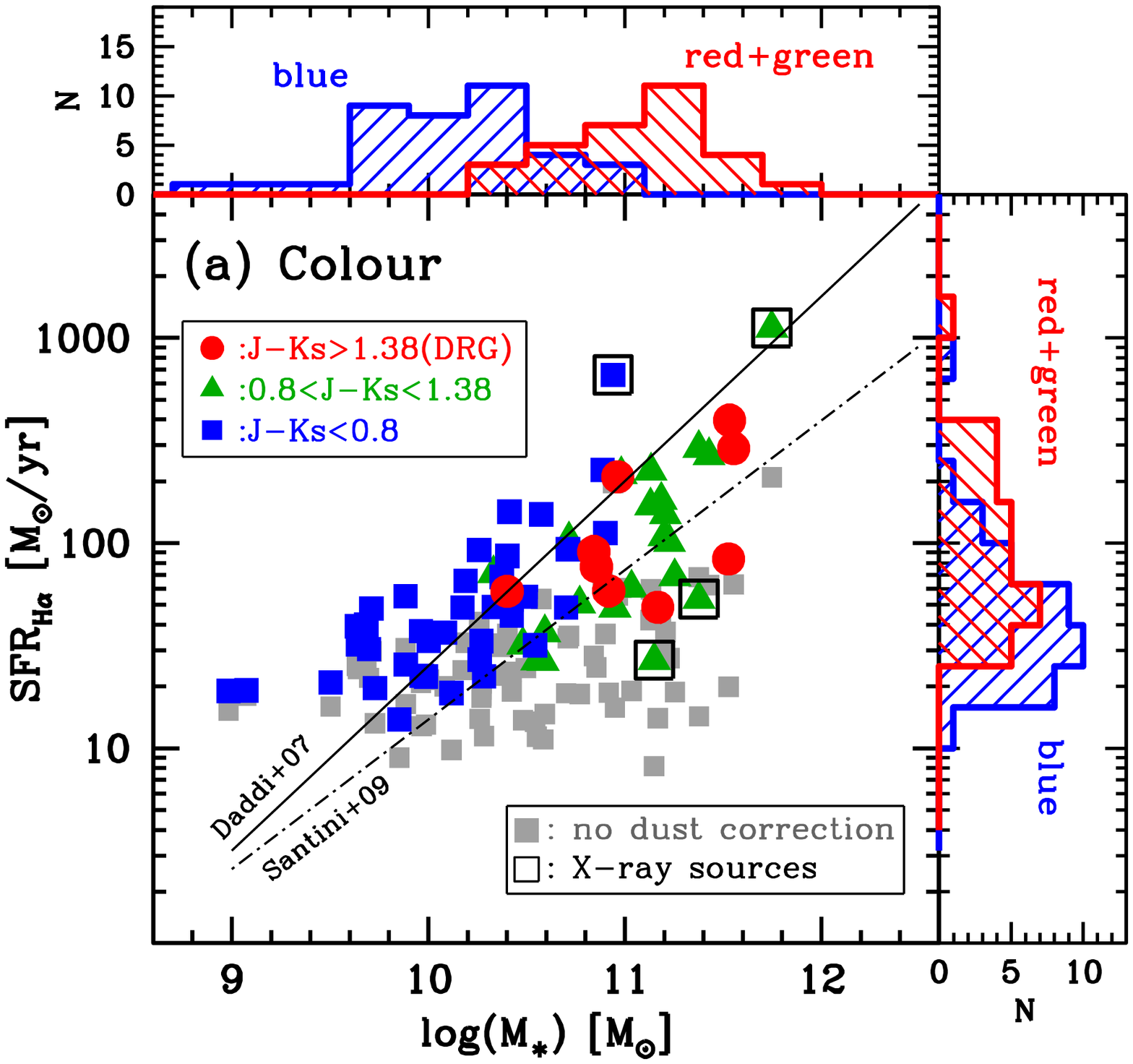}
    \hspace{3mm}
    \epsfxsize 0.48\hsize
    \epsfbox{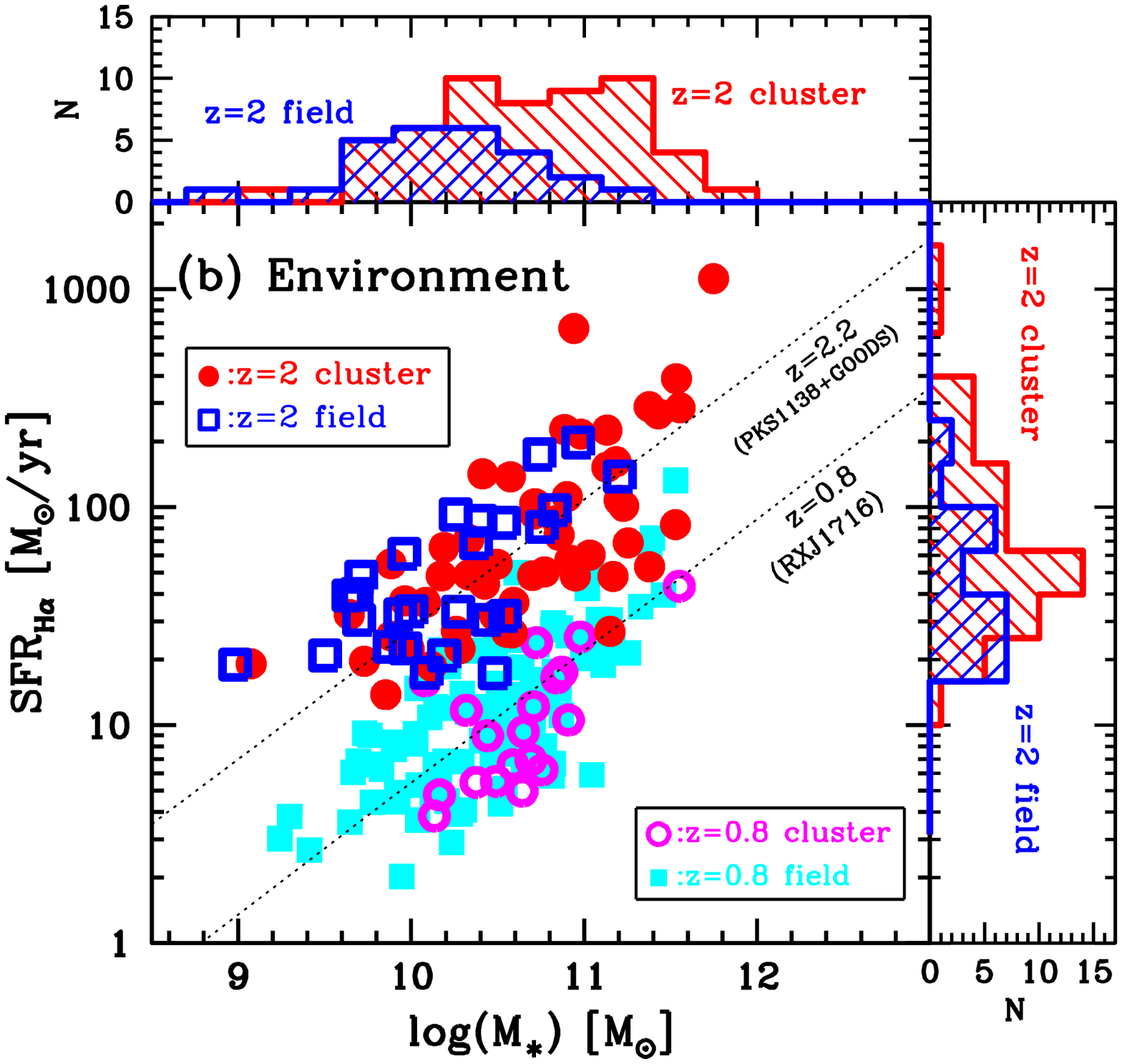}
   \end{center} 
   \vspace{-4mm}
   \caption{(a): Star formation rates of H$\alpha$ emitters plotted against their stellar masses. The colour symbols show the different $(J-K_s)$ colours as indicated. The grey points are in the case of no dust extinction correction, and the open squares indicate X-ray detected emitters. We also show the ``main sequence'' of $z\sim2$ galaxies from Daddi et al.\ (2007) and Santini et al.\ (2009). The histograms in the top and right panels show the $M_*$ and SFR distribution of each colour population. Note that we excluded the radio galaxy in this plot due to the large uncertainty in deriving its $M_*$ and SFR. (b): The same plot but changing the symbol type based on the environment. Note that the field galaxies include the southern half of the PKS\,1138 and GOODS-N field from Tadaki et al.\ (2011). We also show the H$\alpha$-selected galaxies in the $z=0.8$ cluster from Koyama et al.\ (2010) as a comparison, applying an average flux correction ($\times$1.4) taking their NB filter incompleteness into account (see Koyama et al.\ 2010). The dotted lines show the best-fit main sequence for our $z=2.2$ and $z=0.8$ data, calculated using galaxies with $M_*>10^{10}M_{\odot}$.   
 }
\label{fig:SFR_vs_Mstar}
 \end{figure*}
%------------------------------------------------------------------------

\subsection{HST morphologies}

We check the {\it Hubble Space Telescope} ({\it HST}\,) morphologies of the H$\alpha$ emitters using the ACS $I_{814}$-band image. Studying morphologies of galaxies within proto-cluster environments would be an interesting approach to understand how the morphology--density relation seen in the local Universe has been built-up, but it is generally not easy due to e.g.\ the difficulty of determining cluster membership (see pioneering works by e.g.\ \citealt{pet07}; \citealt{pap12}; \citealt{zir12}). We use the pipeline reduced data, which have been retrieved from the {\it Hubble} Legacy Archive, as used in \cite{tan10b}. Note that the $I_{814}$-band traces the rest-frame NUV light for $z\sim 2$ galaxies, and therefore it does not necessarily represent their stellar mass distribution. However, our motivation here is to understand whether the rest-frame UV light is centrally concentrated (dominated by a point source) or spatially extended, which could also be a crude test of the presence of AGNs. The {\it HST} data covers 5.5$'\times$\,3.5$'$ around the PKS\,1138, and 54 emitters are located within this {\it HST} data coverage (but one of them is on a bleed from a bright star). In Fig.~\ref{fig:HST_snapshot}, we show the postage stamps of the available $I_{814}$-band morphologies of our ($K_s$-detected) H$\alpha$ emitters on the colour versus stellar-mass plane. Fig.~\ref{fig:HST_snapshot} clearly shows that a large fraction of the H$\alpha$ emitters display irregular/clumpy morphologies with spatially extended diffuse light, and this trend may be stronger for massive red/green emitters. We attempt to quantify this trend by calculating the "Gini" coefficient ({\it G}) of the H$\alpha$ emitters, using the selected pixels in the segmentation map created by {\sc SExtractor} (we use the pixels within 3$''\times$3$''$ region around each source to minimize the effect of nearby sources). In Fig.~\ref{fig:HST_snapshot}, we plot the {\it G} value as a function of stellar mass, and find that there is a weak trend that red/green massive emitters tend to have smaller {\it G}, hence more uniform light distributions. We caution that the low-mass blue galaxies may be (intrinsically) too small to be resolved, which may produce larger {\it G} values. Also, the smaller {\it G} values for red/green galaxies may be because the central, nucleated starburst activity is hidden by dust (but we need a direct identification of this using e.g.\ high-resolution sub-millimeter observation with ALMA). 

The clumpy morphologies may not be surprising because the rest-frame UV morphologies tend to be biased to highly star-forming regions. Nevertheless, these clumpy morphologies would suggest that the presence of extended star-forming regions within those galaxies, and therefore it is unlikely that a majority of the H$\alpha$ emitters are dominated by pure AGNs. In fact, the X-ray detected H$\alpha$ emitters (ID=5, 82, 83, except for the central radio galaxy) show relatively compact morphologies compared to the others (and tend to have larger {\it G} values; see open squares in the top-right panel in Fig.~\ref{fig:HST_snapshot}). Although galaxies with such compact UV morphologies are not the dominant population in our sample, there are a few more galaxies with large {\it G} values at the high mass end. These galaxies may be additional AGN candidates, but it is impossible to discriminate between nucleated starbursts and AGNs at this moment. We can also notice that the clumpy morphologies may be most prominent in green emitters. Most of the green emitters (except for the X-ray detected ones) have complex morphologies, and many of them seem to accompany multiple cores and/or extended diffuse light.  This could suggest that their activity are triggered by merger events, but firm conclusion require rest-frame optical morphologies, which will be less biased to star-forming regions. Note that a fraction of massive red/green emitters are very faint in the {\it HST} image (see e.g.\ ID\,52, 50, 44, 26). We can still see some diffuse UV emission from these galaxies, and so we expect that their UV light are heavily obscured by dust.  Indeed, a recent MIR spectroscopic study by \cite{ogl12} shows prominent PAH emissions in the rest-frame MIR spectra of ID\,44 and 52 (corresponding to their HAE\,131 and HAE\,229, respectively), which is another evidence that these galaxies have dusty star formation within those galaxies.

\subsection{Star formation "Main Sequence" in cluster environment}

\subsubsection{Colour dependence on the SFR vs. $M_*$ plane}

Recent studies have suggested a  correlation between stellar mass and SFRs of distant star-forming galaxies in the sense that massive galaxies tend to have higher SFRs (e.g.\ \citealt{elb07}; \citealt{dad07}; \citealt{kaj10}), the so-called ``Main Sequence'' of star-forming galaxies. Using the large sample of star-forming galaxies in a proto-cluster environment from our study, we  test if the proto-cluster galaxies are on the same correlation as reported for field galaxies at $z\sim 2$. We plot in Fig.~\ref{fig:SFR_vs_Mstar}a the SFRs of H$\alpha$ emitters in the PKS\,1138 field against their stellar mass. We also show the main sequence of $z\sim 2$ galaxies reported in the literature (\citealt{dad07}; \citealt{san09}) for comparison. Although the main sequence of field galaxies are slightly different between the various literature studies (depending on the sample selection or derivation of physical quantities), our H$\alpha$ emitters are broadly consistent with the literature work. We find that the significant outliers tend to be the X-ray detected sources at the massive end. For these galaxies, AGNs will contribute to the total H$\alpha$ emission and/or continuum level, so that their SFRs and/or stellar masses may not be accurate. It is clear that the redder H$\alpha$ emitters tend to dominate the massive end of the main sequence (i.e.\ higher SFR and masses than blue population). Although we cannot see a significant difference in SFR between red and blue population, it is possible that the red emitters have even higher SFRs, because we may be still underestimating the dust extinction correction in the most dusty systems, detected at the high mass end (see \S4.4.4). A caveat on this plot is that the lack of low-mass red sources is partly due to the $J$-band flux limit (see the colour--magnitude diagram in Fig.~\ref{fig:map_all}), but the lack of blue massive galaxies would be a real phenomenon. 

\subsubsection{Environmental dependence on the SFR vs. $M_*$ plane}

In Fig.~\ref{fig:SFR_vs_Mstar}b, we show the  SFR versus $M_*$ plot, but this time splitting the sample into two  based on their environment. Again, we add the GOODS-N H$\alpha$ emitters from \cite{tad11} to the low-density sample to improve the statistics. We find that the galaxies in the high- and low-density environments show the same main sequence, but with the massive end of the main sequence being dominated by galaxies from the higher-density regions of the proto-cluster. The histograms show the $M_*$ distribution for each environment (i.e.\ stellar mass function of star-forming galaxies), confirming the trend that massive galaxies are more numerous in the higher-density proto-cluster environment. A KS test shows that the probability that these two distributions are from the same parent population is only 0.5\%. We also show the SFR distribution in each environment in the right-hand panel of Fig.~\ref{fig:SFR_vs_Mstar}b, showing a possible trend that the cluster galaxies have higher SFRs compared to field galaxies at $z\sim 2$. This would be a natural consequence of the fact that the massive emitters are more numerous in the cluster environment, and again, the trend would be even stronger because we may be underestimating the extinction correction for these dusty galaxies (see discussion in \S4.4.4). Finally, we note that there are two sources located significantly above the main sequence in our $z\sim 2$ cluster sample, but these galaxies are X-ray detected AGNs (as mentioned in \S4.4.1). Again, our conclusion does not change even if we exclude all the X-ray detected sources. Therefore we conclude that there is no detectable difference between the main sequence in the proto-cluster and general field environment, while the stellar mass distribution within the sequence shows a significant environmental dependence.  

\subsubsection{The evolving main sequence in distant cluster environment}

It is also interesting to test the redshift evolution of the main sequence in cluster environments. In Fig.\ref{fig:SFR_vs_Mstar}b, we plot the distribution of the SFR and stellar masses for H$\alpha$-selected galaxies in a $z\sim 0.8$ cluster from \cite{koy10}. We see a significant difference in the main sequence between $z\sim 0.8$ and $z\sim 2.2$ sample (a factor of $\sim$\,5), which is roughly consistent with the specific SFR decline for field galaxies reported in the literature (e.g.\ \citealt{kar11}; \citealt{whi12}). We note that the $z\sim 0.8$ H$\alpha$ study by \cite{koy10} covers a wide range in environment from the very rich cluster core to its surrounding field, which allows us to split the sample into two environmental bins to test the environmental dependence at $z\sim 0.8$. Fig.~\ref{fig:SFR_vs_Mstar}b shows that the galaxies at $z\sim 0.8$ in high- and low-density environments galaxies broadly follow the same general sequence, but there may be some interesting hints on the galaxy evolution in different environment. 

Firstly, we notice that the massive star-forming galaxies with $M_*\gsim 10^{11}M_{\odot}$ are not present in the $z\sim 0.8$ cluster. We recall that the massive end of the main sequence is dominated by galaxies in high-density environment at $z\sim 2$, but the massive end of the $z\sim 0.8$ main sequence is {\it not} dominated by galaxies in high-density environments. This result would suggest that star-forming activity of the massive star-forming population found in $z\sim 2$ proto-cluster environments have been quenched over the intervening $\sim$\,3 Gyrs ($1\lsim z \lsim 2$), so that they cannot be identified as star-forming at $z\sim 0.8$ any more, confirming the ``down-sizing'' evolution of galaxies in cluster environment (e.g.\ \citealt{tan05}). 

Another interesting hint we can obtain from Fig.~\ref{fig:SFR_vs_Mstar} is that there seems to be a small offset (a factor of $\sim$\,1.5--2) between the main sequence of high-density and low-density galaxies at $z\sim 0.8$. We find that $>$\,80\% of the $z\sim 0.8$ high-density sample are located below the best-fit relation calculated for the whole $z\sim 0.8$ sample (see Fig.~\ref{fig:SFR_vs_Mstar}b). Although the significance is not great (1.7-$\sigma$ level), this may be an interesting implication that galaxies have evolved more rapidly in high-density, cluster environments than in the lower-density field environment at $1\lsim z \lsim 2$ (see also \citealt{tad11}). In fact, the lower specific SFR of galaxies in high-density environments is consistent with some literature works (e.g. \citealt{pat09}; \citealt{vul10}), but the environmental dependence of star-forming activity at $z\sim 1$ is still under debate (e.g.\ \citealt{elb07}; \citealt{muz12}). We caution that the definition of environment is not exactly the same between our $z\sim 0.8$ and $z\sim 2.2$ samples (we divide the $z\sim 0.8$ sample into cluster/field environment at $\log\Sigma_5=2$ using the local number density of photo-$z$ selected galaxies; see \citealt{koy07}; 2008). Also, our result is based on just one cluster at each redshift, and that these two clusters may not be necessarily on the same evolutionary track (i.e.\ it is not clear if PKS\,1138 is really the progenitor of our $z\sim 0.8$ cluster). It is clearly important to verify this trend based using larger, unbiased sample of galaxies at different redshifts, but Fig.~\ref{fig:SFR_vs_Mstar} may be giving us a hint of the role of environment in the evolution of galaxies at high redshifts.

\subsubsection{A caveat on the effect of dust extinction correction}

Our survey suggests that $M_*$ distribution of star-forming galaxies is likely to depend on the environment at $z\sim 2$, in the sense that massive star-forming galaxies are more numerous in  high-density environments. This result is qualitatively consistent with pioneering work which investigated the environmental dependence of galaxy properties at $z\sim 2$--3 (e.g.\ \citealt{ste05}; \citealt{hat11}; \citealt{mat11}). We also obtained an interesting hint that the $z\sim 2$ galaxies in high-density regions have higher SFR than  galaxies in lower-density environments at the same redshift, but the SFR-related result should be treated with caution because it may be affected by the uncertainty in the dust extinction correction. We recall that we adopted a stellar-mass dependent extinction correction for the H$\alpha$ luminosities (see \S3.3.2). Varying this correction has little impact on the stellar mass measurements, but we here discuss some possible effects on the SFR-related results. 

As a check of the influence of the reddening corrections, we apply (1) the conventional 1-magnitude correction for dust reddening in H$\alpha$ samples and (2) the SFR-dependent correction presented by \cite{gar10a}. We find that these methods both make the main sequence slope flatter, so that the environmental dependence in SFR is likely to be washed out. This is in fact consistent with some earlier works which reported no environmental variations in SFR distribution (i.e.\ the shape of H$\alpha$ luminosity functions rather than its normalisation or characteristic luminosity) among star-forming galaxies at $z\sim 2$ (\citealt{hat11}; \citealt{mat11}; see also similar results at $z\sim 1$ by \citealt{koy10}; \citealt{sob11}). However, as far as we believe the main sequence reported in the literature, the constant 1-magnitude correction tends to underestimate the extinction of massive galaxies, indeed we estimate the extinction of such massive H$\alpha$ emitters to be $A_{\rm H\alpha}\simeq$\,2--3 mag based on the MIR photometry. This is even higher than that estimated using the $M_*$-dependent correction we adopt in this study, although the SFR derived from rest-frame 8$\mu$m photometry might be overestimated (e.g.\ \citealt{nor10}). On the other hand, the SFR-dependent extinction correction seems to be overestimating the extinction of low-mass galaxies (again, if we assume that they are the normal, main-sequence galaxies). Therefore, we expect that the $M_*$-dependent correction adopted in this study would be more realistic at this stage, although it is currently impossible to accurately measure the extinction effect on individual galaxies.

%%%%%%%%%%%%%%%%%%%%%%%%%%%%%%%%%%%%%%%%%%%%%%%%%%%%%%%%%%%%%%%%%%%%%%%%%%
% SUMMARY & CONCLUSIONS
%%%%%%%%%%%%%%%%%%%%%%%%%%%%%%%%%%%%%%%%%%%%%%%%%%%%%%%%%%%%%%%%%%%%%%%%%%
\section{Summary and Conclusions}
\label{sec:summary}

We have presented a panoramic survey for H$\alpha$ emitters in the proto-cluster around PKS\,1138 at $z=2.16$ using a narrow-band  filter with MOIRCS on Subaru. We study properties of star-forming galaxies in the proto-cluster, and discuss their environmental dependence. Our results are summarized as follows. 

(1) A prominent large-scale structure is identified around PKS\,1138. We find that the PKS\,1138 proto-cluster is embedded in a $\gsim$\,10 Mpc-scale  filament traced by H$\alpha$ emitters and extending from north-east to south-west through the radio galaxy. We also find a possible filament running towards south-east from the radio galaxy, suggesting that the PKS\,1138 field is located at a node of the cosmic web at $z=2.16$. Several emitters are likely to be AGN (five are X-ray detected), but we expect that the majority of our sample are star-forming galaxies, based on the SED analysis and on their clumpy/extended rest-frame UV morphologies in {\it HST} imaging.  

(2) Galaxies in the $z=2$ proto-cluster environment tend to have redder colours and higher stellar masses (and probably larger SFRs) compared to those in underdense regions. In particular, we find that a substantial fraction of H$\alpha$ emitters in PKS\,1138 show red colours with $(J-K_s)\gsim1$, and in some extreme cases, they satisfy the DRG criteria. Such red H$\alpha$ emitters tend to have very large stellar masses ($M_*\gsim$\,10$^{11}M_{\odot}$), compared to the more normal blue galaxies. This is an evidence that while many cluster galaxies are still in a vigorously star-forming phase, they have already formed a large part of their stellar mass by $z\sim 2$.

(3) The red H$\alpha$ emitters tend to be detected in the {\it Spitzer} MIPS 24$\mu$m imaging, suggesting they are luminous and dusty. Within the MIPS data coverage, we find that 14 out of 29 (48\%) are detected at 24$\mu$m from our photometrically-defined sample of red/green emitters, while only 1 out of 24 (4\%) is detected for the blue emitter sample. Such 24$\mu$m-detected H$\alpha$ emitters are estimated to have SFR$_{\rm IR}\gsim$\,100\,M$_{\odot}$\,yr$^{-1}$ (i.e.\ ULIRG class) and heavy dust extinction with $A_{\rm{H}\alpha}\sim2$--3 mag. Therefore we suggest that these intense starbursts are a common population in the young proto-cluster environment, while they are absent from the cores of lower redshift cluster at $z\lsim 1$. 

(4) The red H$\alpha$ emitters may be a cluster phenomenon. Our survey reveals that the red H$\alpha$ emitters are preferentially found in the higher-density proto-cluster region, while the blue H$\alpha$ emitters are more widely spread. Therefore, we expect that the red, massive star-forming galaxies are the key population driven by environmental effects in the early Universe, which go onto become present-day passive cluster galaxies. 

(5) The H$\alpha$ emitters in proto-cluster environments are  on broadly the same ``main sequence'' of $z\sim 2$ star-forming galaxies as those in the field, but there is an excess of massive systems compared to lower-density environments. In contrast, this trend may not be present at $z\sim 1$, where the most massive galaxies in high-density regions are not forming stars any more. We also find a tentative hint that the star-forming galaxies in cluster environments at $z\sim 0.8$  may have lower star-formation activity than in field  at the same redshift, implying that the accelerated galaxy evolution we see is associated with the cluster environment at $1\lsim z \lsim 2$.

%%%%%%%%%%%%%%%%%%%%%%%%%%%%%%%%%%%%%%%%%%%%%%%%%%%%%%%%%%%%%%%%%%%%%%
% ACKNOWLEDGEMENT
%%%%%%%%%%%%%%%%%%%%%%%%%%%%%%%%%%%%%%%%%%%%%%%%%%%%%%%%%%%%%%%%%%%%%%
\section*{Acknowledgment}
We thank the anonymous referee for their constructive and helpful comments. The optical and NIR data newly presented in this paper were collected at the Subaru Telescope, which is operated by the National Astronomical Observatory of Japan (NAOJ). The MIR data used in this paper is taken with the {\it Spitzer Space Telescope}, which is operated by the Jet Propulsion Laboratory, California Institute of Technology under a contract with NASA. We thank Dr.\ David Sobral for providing us with the HiZELS catalogue. This work was financially supported by a Grant-in-Aid for the Scientific Research (Nos.\, 21340045; 23740144; 24244015) by the Japanese Ministry of Education, Culture, Sports and Science. This work is in part supported by World Premier International Research Center Initiative (WPI Initiative), MEXT, Japan. Y.K. and K.T. acknowledge support from the Japan Society for the Promotion of Science (JSPS) through JSPS research fellowships for Young Scientists. I.R.S. acknowledges support from a Leverhulme Fellowship and from STFC. J.K. acknowledges support from the German Science Foundation (DFG) via German-Israeli Project Cooperation grant STE1869/1-1.GE625/15-1.

%------------------------------------------------------------------

%------------------------------------------------------------------


\begin{thebibliography}{99}
\bibitem[\protect\citeauthoryear{Andreon et al.}{2009}]{and09} 
Andreon S., Maughan B., Trinchieri G., Kurk J., 2009, A\&A, 507, 147 
\bibitem[\protect\citeauthoryear{Balogh et al.}{2002}]{bal02} 
Balogh M.~L., Couch W.~J., Smail I., Bower R.~G., Glazebrook K., 2002, 
MNRAS, 335, 10 
\bibitem[\protect\citeauthoryear{Bauer et al.}{2011}]{bau11} 
Bauer A.~E., Gr{\"u}tzbauch R., J{\o}rgensen I., Varela J., Bergmann M., 
2011, MNRAS, 411, 2009 
\bibitem[\protect\citeauthoryear{Bertin \& Arnouts}{1996}]{ber96}
Bertin E., Arnouts S., 1996, A\&AS, 117, 393 
\bibitem[\protect\citeauthoryear{Bower, Lucey, \& Ellis}{1992}]{bow92}
Bower R.~G., Lucey J.~R., Ellis R.~S., 1992, MNRAS, 254, 601 
\bibitem[\protect\citeauthoryear{Bruzual \& Charlot}{2003}]{bc03} 
Bruzual G., Charlot S., 2003, MNRAS, 344, 1000 
\bibitem[\protect\citeauthoryear{Bunker et al.}{1995}]{bun95} 
Bunker A.~J., Warren S.~J., Hewett P.~C., Clements D.~L., 1995, MNRAS, 273, 
513 
\bibitem[\protect\citeauthoryear{Butcher \& Oemler}{1984}]{but84} 
Butcher H., Oemler A., Jr., 1984, ApJ, 285, 426
\bibitem[\protect\citeauthoryear{Cardelli, Clayton, \& Mathis}{1989}]{car89} 
Cardelli J.~A., Clayton G.~C., Mathis J.~S., 1989, ApJ, 345, 245 
\bibitem[\protect\citeauthoryear{Carilli et al.}{2002}]{car02} 
Carilli C.~L., Harris D.~E., Pentericci L., R{\"o}ttgering H.~J.~A., Miley G.~K., Kurk J.~D., van Breugel W., 2002, ApJ, 567, 781 
\bibitem[\protect\citeauthoryear{Carilli et al.}{1997}]{car97} 
Carilli C.~L., Roettgering H.~J.~A., van Ojik R., Miley G.~K., van Breugel W.~J.~M., 1997, ApJS, 109, 1 
\bibitem[\protect\citeauthoryear{Chabrier}{2003}]{cha03} 
Chabrier G., 2003, PASP, 115, 763 
\bibitem[\protect\citeauthoryear{Cooper et al.}{2008}]{coo08} 
Cooper M.~C., et al., 2008, MNRAS, 383, 1058 
\bibitem[\protect\citeauthoryear{Couch et al.}{2001}]{cou01} 
Couch W.~J., Balogh M.~L., Bower R.~G., Smail I., Glazebrook K., Taylor M., 
2001, ApJ, 549, 820 
\bibitem[\protect\citeauthoryear{Croft et al.}{2005}]{cro05} 
Croft S., Kurk J., van Breugel W., Stanford S.~A., de Vries W., Pentericci 
L., R{\"o}ttgering H., 2005, AJ, 130, 867 
\bibitem[\protect\citeauthoryear{Daddi et al.}{2007}]{dad07} 
Daddi E., et al., 2007, ApJ, 670, 156 
\bibitem[\protect\citeauthoryear{Daddi et al.}{2004}]{dad04} 
Daddi E., Cimatti A., Renzini A., Fontana A., Mignoli M., Pozzetti L., 
Tozzi P., Zamorani G., 2004, ApJ, 617, 746 
\bibitem[\protect\citeauthoryear{Doherty et al.}{2010}]{doh10} 
Doherty M., et al., 2010, A\&A, 509, A83 
\bibitem[\protect\citeauthoryear{Dressler}{1980}]{dre80} 
Dressler A., 1980, ApJ, 236, 351 
\bibitem[\protect\citeauthoryear{Eastman et al.}{2007}]{eas07} 
Eastman J., Martini P., Sivakoff G., Kelson D.~D., Mulchaey J.~S., Tran K.-V., 2007, ApJ, 664, L9 
\bibitem[\protect\citeauthoryear{Elbaz et al.}{2007}]{elb07} 
Elbaz D., et al., 2007, A\&A, 468, 33 
\bibitem[\protect\citeauthoryear{Fassbender et al.}{2011}]{fas11} 
Fassbender R., et al., 2011, A\&A, 527, L10 
\bibitem[\protect\citeauthoryear{Franx et al.}{2003}]{fra03} 
Franx M., et al., 2003, ApJ, 587, L79 
\bibitem[\protect\citeauthoryear{Garn \& Best}{2010}]{gar10b} 
Garn T., Best P.~N., 2010, MNRAS, 409, 421 
\bibitem[\protect\citeauthoryear{Garn et al.}{2010}]{gar10a} 
Garn T., et al., 2010, MNRAS, 402, 2017 
\bibitem[\protect\citeauthoryear{Geach et al.}{2008}]{gea08} 
Geach J.~E., Smail I., Best P.~N., Kurk J., Casali M., Ivison R.~J., Coppin 
K., 2008, MNRAS, 388, 1473 
\bibitem[\protect\citeauthoryear{Geach et al.}{2006}]{gea06} 
Geach J.~E., et al., 2006, ApJ, 649, 661 
\bibitem[\protect\citeauthoryear{Gobat et al.}{2011}]{gob11} 
Gobat R., et al., 2011, A\&A, 526, A133 
\bibitem[\protect\citeauthoryear{G{\'o}mez et al.}{2003}]{gom03} 
G{\'o}mez P.~L., et al., 2003, ApJ, 584, 210 
\bibitem[\protect\citeauthoryear{Goto et al.}{2003}]{got03} 
Goto T., Yamauchi C., Fujita Y., Okamura S., Sekiguchi M., Smail I., Bernardi M., Gomez P.~L., 2003, MNRAS, 346, 601 
\bibitem[\protect\citeauthoryear{Haines et al.}{2009}]{hai09} 
Haines C.~P., et al., 2009, ApJ, 704, 126
\bibitem[\protect\citeauthoryear{Hatch et al.}{2011}]{hat11} 
Hatch N.~A., Kurk J.~D., Pentericci L., Venemans B.~P., Kuiper E., Miley 
G.~K., R{\"o}ttgering H.~J.~A., 2011, MNRAS, 415, 2993 
\bibitem[\protect\citeauthoryear{Hatch et al.}{2009}]{hat09} 
Hatch N.~A., Overzier R.~A., Kurk J.~D., Miley G.~K., R{\"o}ttgering 
H.~J.~A., Zirm A.~W., 2009, MNRAS, 395, 114 
\bibitem[\protect\citeauthoryear{Hatch et al.}{2008}]{hat08} 
Hatch N.~A., Overzier R.~A., R{\"o}ttgering H.~J.~A., Kurk J.~D., Miley 
G.~K., 2008, MNRAS, 383, 931
\bibitem[\protect\citeauthoryear{Hayashi et al.}{2012}]{hay12} 
Hayashi M., Kodama T., Tadaki K.-i., Koyama Y., Tanaka I., 2012, ApJ, 757, 15 
\bibitem[\protect\citeauthoryear{Hayashi et al.}{2011}]{hay11} 
Hayashi M., Kodama T., Koyama Y., Tadaki K.-I., Tanaka I., 2011, MNRAS, 415, 2670 
\bibitem[\protect\citeauthoryear{Hayashi et al.}{2010}]{hay10} 
Hayashi M., Kodama T., Koyama Y., Tanaka I., Shimasaku K., Okamura S., 2010, MNRAS, 402, 1980 
\bibitem[\protect\citeauthoryear{Hilton et al.}{2010}]{hil10} 
Hilton M., et al., 2010, ApJ, 718, 133 
\bibitem[\protect\citeauthoryear{Hopkins \& Beacom}{2006}]{hop06} 
Hopkins A.~M., Beacom J.~F., 2006, ApJ, 651, 142 
\bibitem[\protect\citeauthoryear{Ichikawa et al.}{2006}]{ich06} 
Ichikawa T., et al., 2006, SPIE, 6269,  
\bibitem[\protect\citeauthoryear{Iye et al.}{2004}]{iye04}
Iye M. et al.\ 2004, PASJ, 56, 381
\bibitem[\protect\citeauthoryear{Kajisawa et al.}{2010}]{kaj10} 
Kajisawa M., Ichikawa T., Yamada T., Uchimoto Y.~K., Yoshikawa T., 
Akiyama M., Onodera M., 2010, ApJ, 723, 129 
\bibitem[\protect\citeauthoryear{Kajisawa et al.}{2006}]{kaj06} 
Kajisawa M., Kodama T., Tanaka I., Yamada T., Bower R., 2006, MNRAS, 371, 577 
\bibitem[\protect\citeauthoryear{Karim et al.}{2011}]{kar11} 
Karim A., et al., 2011, ApJ, 730, 61 
\bibitem[\protect\citeauthoryear{Kennicutt}{1998}]{ken98} 
Kennicutt R.~C., Jr., 1998, ARA\&A, 36, 189 
\bibitem[\protect\citeauthoryear{Kodama et al.}{2012}]{kod12} 
Kodama T., et al., 2012, in prep.
\bibitem[\protect\citeauthoryear{Kodama et al.}{2007}]{kod07} 
Kodama T., Tanaka I., Kajisawa M., Kurk J., Venemans B., De Breuck C., 
Vernet J., Lidman C., 2007, MNRAS, 377, 1717 
\bibitem[\protect\citeauthoryear{Kodama et al.}{2004}]{kod04} 
Kodama T., Balogh M.~L., Smail I., Bower R.~G., Nakata F., 2004, MNRAS, 
354, 1103 
\bibitem[\protect\citeauthoryear{Kodama et al.}{2001}]{kod01} 
Kodama T., Smail I., Nakata F., Okamura S., Bower R.~G., 2001, ApJ, 562, L9 
\bibitem[\protect\citeauthoryear{Kodama, Bell, \& Bower}{1999}]{kod99}
Kodama T., Bell E.~F., Bower R.~G., 1999, MNRAS, 302, 152
\bibitem[\protect\citeauthoryear{Kodama et al.}{1998}]{kod98} 
Kodama T., Arimoto N., Barger A.~J., Arag'on-Salamanca A., 1998, A\&A, 334, 99 
\bibitem[\protect\citeauthoryear{Kodama \& Arimoto}{1997}]{kod97} 
Kodama T., Arimoto N., 1997, A\&A, 320, 41 
\bibitem[\protect\citeauthoryear{Koyama et al.}{2011}]{koy11} 
Koyama Y., Kodama T., Nakata F., Shimasaku K., Okamura S., 2011, ApJ, 734, 
66 
\bibitem[\protect\citeauthoryear{Koyama et al.}{2010}]{koy10} 
Koyama Y., Kodama T., Shimasaku K., Hayashi M., Okamura S., Tanaka I., 
Tokoku C., 2010, MNRAS, 403, 1611 
\bibitem[\protect\citeauthoryear{Koyama et al.}{2008}]{koy08} 
Koyama Y., et al., 2008, MNRAS, 391, 1758 
\bibitem[\protect\citeauthoryear{Koyama et al.}{2007}]{koy07} 
Koyama Y., Kodama T., Tanaka M., Shimasaku K., Okamura S., 2007, MNRAS, 
382, 1719 
\bibitem[\protect\citeauthoryear{Kroupa}{2001}]{kro01} 
Kroupa P., 2001, MNRAS, 322, 231 
\bibitem[\protect\citeauthoryear{Kuiper et al.}{2011}]{kui11} 
Kuiper E., et al., 2011, MNRAS, 415, 2245 
\bibitem[\protect\citeauthoryear{Kurk et al.}{2004b}]{kur04b} 
Kurk J.~D., Pentericci L., Overzier R.~A., R{\"o}ttgering H.~J.~A., Miley G.~K., 2004, A\&A, 428, 817 
\bibitem[\protect\citeauthoryear{Kurk et al.}{2004a}]{kur04a} 
Kurk J.~D., Pentericci L., R{\"o}ttgering H.~J.~A., Miley G.~K., 2004, A\&A, 428, 793 
\bibitem[\protect\citeauthoryear{Kurk et al.}{2000}]{kur00} 
Kurk J.~D., et al., 2000, A\&A, 358, L1 
\bibitem[\protect\citeauthoryear{Lagache, Dole, \& Puget}{2003}]{lag03} 
Lagache G., Dole H., Puget J.-L., 2003, MNRAS, 338, 555 
\bibitem[\protect\citeauthoryear{Lewis et al.}{2002}]{lew02} 
Lewis I., et al., 2002, MNRAS, 334, 673 
\bibitem[\protect\citeauthoryear{Marcillac et al.}{2007}]{mar07} 
Marcillac D., Rigby J.~R., Rieke G.~H., Kelly D.~M., 2007, ApJ, 654, 825 
\bibitem[\protect\citeauthoryear{Makovoz \& Khan}{2005}]{mak05} 
Makovoz D., Khan I., 2005, ASPC, 347, 81 
\bibitem[\protect\citeauthoryear{Martini, Sivakoff, \& Mulchaey}{2009}]{mar09} 
Martini P., Sivakoff G.~R., Mulchaey J.~S., 2009, ApJ, 701, 66 
\bibitem[\protect\citeauthoryear{Matsuda et al.}{2011}]{mat11} 
Matsuda Y., et al., 2011, MNRAS, 416, 2041 
\bibitem[\protect\citeauthoryear{Mayo et al.}{2012}]{may12} 
Mayo J.~H., Vernet J., De Breuck C., Galametz A., Seymour N., Stern D., 2012, A\&A, 539, A33 
\bibitem[\protect\citeauthoryear{Miley \& De Breuck}{2008}]{mil08} 
Miley G., De Breuck C., 2008, A\&ARv, 15, 67 
\bibitem[\protect\citeauthoryear{Miley et al.}{2006}]{mil06} 
Miley G.~K., et al., 2006, ApJ, 650, L29 
\bibitem[\protect\citeauthoryear{Miyazaki et al.}{2002}]{miy02}
Miyazaki S., et al., 2002, PASJ, 54, 833 
\bibitem[\protect\citeauthoryear{Muzzin et al.}{2012}]{muz12} 
Muzzin A., et al., 2012, ApJ, 746, 188 
\bibitem[\protect\citeauthoryear{Nordon et al.}{2010}]{nor10} 
Nordon R., et al., 2010, A\&A, 518, L24 
\bibitem[\protect\citeauthoryear{Ogle et al.}{2012}]{ogl12} 
Ogle P., Davies J.~E., Appleton P.~N., Bertincourt B., Seymour N., Helou 
G., 2012, ApJ, 751, 13 
\bibitem[\protect\citeauthoryear{Ouchi et al.}{2004}]{ouc04} 
Ouchi M., et al., 2004, ApJ, 611, 660
\bibitem[\protect\citeauthoryear{Overzier et al.}{2006}]{ove06} 
Overzier R.~A., et al., 2006, ApJ, 637, 58 
\bibitem[\protect\citeauthoryear{Papovich et al.}{2012}]{pap12} 
Papovich C., et al., 2012, ApJ, 750, 93 
\bibitem[\protect\citeauthoryear{Papovich et al.}{2010}]{pap10} 
Papovich C., et al., 2010, ApJ, 716, 1503 
\bibitem[\protect\citeauthoryear{Patel et al.}{2009}]{pat09} 
Patel S.~G., Holden B.~P., Kelson D.~D., Illingworth G.~D., Franx M., 2009, 
ApJ, 705, L67 
\bibitem[\protect\citeauthoryear{Pentericci et al.}{2002}]{pen02} 
Pentericci L., Kurk J.~D., Carilli C.~L., Harris D.~E., Miley G.~K., R{\"o}ttgering H.~J.~A., 2002, A\&A, 396, 109 
\bibitem[\protect\citeauthoryear{Pentericci et al.}{2000}]{pen00} 
Pentericci L., et al., 2000, A\&A, 361, L25 
\bibitem[\protect\citeauthoryear{Pentericci et al.}{1998}]{pen98} 
Pentericci L., Roettgering H.~J.~A., Miley G.~K., Spinrad H., McCarthy P.~J., van Breugel W.~J.~M., Macchetto F., 1998, ApJ, 504, 139 
\bibitem[\protect\citeauthoryear{Pentericci et al.}{1997}]{pen97} 
Pentericci L., Roettgering H.~J.~A., Miley G.~K., Carilli C.~L., McCarthy P., 1997, A\&A, 326, 580 
\bibitem[\protect\citeauthoryear{Peter et al.}{2007}]{pet07} 
Peter A.~H.~G., Shapley A.~E., Law D.~R., Steidel C.~C., Erb D.~K., Reddy 
N.~A., Pettini M., 2007, ApJ, 668, 23 
\bibitem[\protect\citeauthoryear{Quadri et al.}{2012}]{qua12} 
Quadri R.~F., Williams R.~J., Franx M., Hildebrandt H., 2012, ApJ, 744, 88 
\bibitem[\protect\citeauthoryear{Raichoor \& Andreon}{2012}]{rai12} 
Raichoor A., Andreon S., 2012, A\&A, 537, A88 
\bibitem[\protect\citeauthoryear{Rieke et al.}{2004}]{rie04} 
Rieke G.~H., et al., 2004, ApJS, 154, 25 
\bibitem[\protect\citeauthoryear{Salpeter}{1955}]{sal55} 
Salpeter E.~E., 1955, ApJ, 121, 161 
\bibitem[\protect\citeauthoryear{Saintonge, Tran, \& Holden}{2008}]{sai08} 
Saintonge A., Tran K.-V.~H., Holden B.~P., 2008, ApJ, 685, L113 
\bibitem[\protect\citeauthoryear{Santini et al.}{2009}]{san09} 
Santini P., et al., 2009, A\&A, 504, 751 
\bibitem[\protect\citeauthoryear{Schlegel, Finkbeiner, \& Davis}{1998}]{sch98} 
Schlegel D.~J., Finkbeiner D.~P., Davis M., 1998, ApJ, 500, 525 
\bibitem[\protect\citeauthoryear{Smail et al.}{1999}]{sma99} 
Smail I., Morrison G., Gray M.~E., Owen F.~N., Ivison R.~J., Kneib J.-P., Ellis R.~S., 1999, ApJ, 525, 609 
\bibitem[\protect\citeauthoryear{Smith et al.}{2012}]{smi12} 
Smith R.~J., Lucey J.~R., Price J., Hudson M.~J., Phillipps S., 2012, 
MNRAS, 419, 3167 
\bibitem[\protect\citeauthoryear{Sobral et al.}{2012}]{sob12} 
Sobral D., Best P.~N., Matsuda Y., Smail I., Geach J.~E., Cirasuolo M., 
2012, MNRAS, 420, 1926 
\bibitem[\protect\citeauthoryear{Sobral et al.}{2011}]{sob11} 
Sobral D., Best P.~N., Smail I., Geach J.~E., Cirasuolo M., Garn T., Dalton 
G.~B., 2011, MNRAS, 411, 675 
\bibitem[\protect\citeauthoryear{Stanford et al.}{2006}]{sta06} 
Stanford S.~A., et al., 2006, ApJ, 646, L13 
\bibitem[\protect\citeauthoryear{Steidel et al.}{2005}]{ste05} 
Steidel C.~C., Adelberger K.~L., Shapley A.~E., Erb D.~K., Reddy N.~A., Pettini M., 2005, ApJ, 626, 44 
\bibitem[\protect\citeauthoryear{Steidel et al.}{2000}]{ste00} 
Steidel C.~C., Adelberger K.~L., Shapley A.~E., Pettini M., Dickinson M., Giavalisco M., 2000, ApJ, 532, 170 
\bibitem[\protect\citeauthoryear{Suzuki et al.}{2008}]{suz08} 
Suzuki R., et al., 2008, PASJ, 60, 1347 
\bibitem[\protect\citeauthoryear{Tadaki et al.}{2012}]{tad12} 
Tadaki K.-i., et al., 2012, MNRAS, 423, 2617 
\bibitem[\protect\citeauthoryear{Tadaki et al.}{2011}]{tad11} 
Tadaki K.-I., Kodama T., Koyama Y., Hayashi M., Tanaka I., Tokoku C., 2011, 
PASJ, 63, 437 
\bibitem[\protect\citeauthoryear{Tanaka et al.}{2011}]{tani11} 
Tanaka I., et al., 2011, PASJ, 63, 415 
\bibitem[\protect\citeauthoryear{Tanaka, Finoguenov, \& Ueda}{2010}]{tan10a} 
Tanaka M., Finoguenov A., Ueda Y., 2010, ApJ, 716, L152 
\bibitem[\protect\citeauthoryear{Tanaka et al.}{2010}]{tan10b} 
Tanaka M., De Breuck C., Venemans B., Kurk J., 2010, A\&A, 518, A18 
\bibitem[\protect\citeauthoryear{Tanaka et al.}{2004}]{tan04} 
Tanaka M., Goto T., Okamura S., Shimasaku K., Brinkmann J., 2004, AJ, 128, 2677 
\bibitem[\protect\citeauthoryear{Tanaka et al.}{2005}]{tan05} 
Tanaka M., Kodama T., Arimoto N., Okamura S., Umetsu K., Shimasaku K., 
Tanaka I., Yamada T., 2005, MNRAS, 362, 268 
\bibitem[\protect\citeauthoryear{Tomczak, Tran, \& Saintonge}{2011}]{tom11} 
Tomczak A.~R., Tran K.-V.~H., Saintonge A., 2011, ApJ, 738, 65 
\bibitem[\protect\citeauthoryear{Tran et al.}{2010}]{tra10} 
Tran K.-V.~H., et al., 2010, ApJ, 719, L126 
\bibitem[\protect\citeauthoryear{Uchimoto et al.}{2008}]{uch08} 
Uchimoto Y.~K., et al., 2008, PASJ, 60, 683 
\bibitem[\protect\citeauthoryear{Venemans et al.}{2007}]{ven07} 
Venemans B.~P., et al., 2007, A\&A, 461, 823 
\bibitem[\protect\citeauthoryear{Villar et al.}{2011}]{vil11} 
Villar V., Gallego J., P{\'e}rez-Gonz{\'a}lez P.~G., Barro G., Zamorano J., 
Noeske K., Koo D.~C., 2011, ApJ, 740, 47 
\bibitem[\protect\citeauthoryear{Vulcani et al.}{2010}]{vul10} 
Vulcani B., Poggianti B.~M., Finn R.~A., Rudnick G., Desai V., Bamford S., 2010, ApJ, 710, L1 
\bibitem[\protect\citeauthoryear{Whitaker et al.}{2012}]{whi12} 
Whitaker K.~E., van Dokkum P.~G., Brammer G., Franx M., 2012, ApJ, 754, L29  
\bibitem[\protect\citeauthoryear{Yagi et al.}{2002}]{yag02} 
Yagi M., Kashikawa N., Sekiguchi M., Doi M., Yasuda N., Shimasaku K., Okamura S., 2002, AJ, 123, 66 
\bibitem[\protect\citeauthoryear{Yamada et al.}{2012}]{yam12} 
Yamada T., Nakamura Y., Matsuda Y., Hayashino T., Yamauchi R., Morimoto N., 
Kousai K., Umemura M., 2012, AJ, 143, 79 
\bibitem[\protect\citeauthoryear{Zirm, Toft, \& Tanaka}{2012}]{zir12} 
Zirm A.~W., Toft S., Tanaka M., 2012, ApJ, 744, 181 
\bibitem[\protect\citeauthoryear{Zirm et al.}{2008}]{zir08} 
Zirm A.~W., et al., 2008, ApJ, 680, 224
\end{thebibliography}
\end{document}